\documentclass{l4dc2027}

\usepackage{times}
\makeatletter
\def\ps@jmlrtps{\let\@mkboth\@gobbletwo\def\@oddhead{}\let\@evenhead\@oddhead
  \def\@oddfoot{\hfill\small\rmfamily\thepage\hfill}\let\@evenfoot\@oddfoot}
\renewcommand*{\@jmlrenddoc}{}
\makeatother
\usepackage{xcolor}
\usepackage{booktabs}
\usepackage{tikz}
\usetikzlibrary{positioning,arrows.meta,fit,backgrounds,calc}

\newcommand{\lsoft}{\ell_{\mathrm{soft}}}
\newcommand{\lcap}{\ell_{\delta}}
\newcommand{\safeset}{\mathcal{S}}
\newcommand{\winspace}{\mathcal{W}}
\newcommand{\relu}{\mathrm{ReLU}}

\title[Reliable Conditioned HJ Reachability]{Reliable Conditioned Hamilton--Jacobi Reachability \\via Subsolution Learning on Exact Local Windows}

\coltauthor{
 \Name{Seunghwan Jang} \Email{seunghwa001@e.ntu.edu.sg}\\
 \Name{Wei Xiao} \Email{wei.xiao@ntu.edu.sg}\\
 \addr Nanyang Technological University, Singapore
}

\begin{document}

\maketitle

\begin{abstract}
Hamilton--Jacobi (HJ) reachability computes the maximal safe set and a safety filter for a given obstacle configuration, but it must be re-solved whenever the obstacles change, which a deployed robot cannot afford.
Learning a single value function conditioned on the obstacle set removes the re-solve, yet it sharpens the central weakness of learned reachability: one network must now cover a whole family of safety problems, and its approximation error carries no preferred sign, so rare obstacle configurations silently produce false-safe states. To address these challenges,
we present a conditioned reachability model designed so that the remaining error falls on the conservative side. Specifically,
our method caps the obstacle margin, which makes the value function provably local, in the sense that obstacles farther than a known threshold cannot affect it, so the model reads only a body-frame window, trains on small scenes, and transfers to scenes of any size with millisecond per-scene preparation.
Moreover, the output structure builds the value bound into the model.
By training with counterexample mining over the finite-dimensional window configuration space, we maximize the claimed safe volume subject to the HJ inequality, so the learned set grows toward the true safe set from the conservative side.
Across all tested obstacle settings, ours is the only method that combines recovery comparable to per-scene DeepReach training with near-zero rollout collisions.
We also extend the method to a 10D quadrotor, again with markedly lower collision rates than the baselines.
\end{abstract}


\section{Introduction}
\label{sec:intro}

\begin{figure}[t]
\centering
\includegraphics[width=\linewidth]
{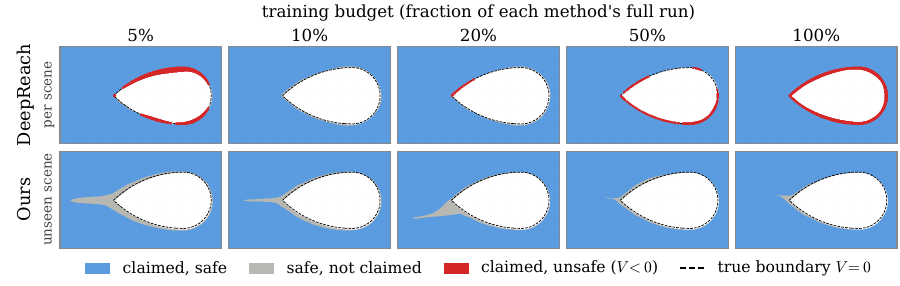}
\vspace{-25px}
\caption{Safe sets claimed by each model across training budgets, compared with the true safe set.
DeepReach errs on either side of the boundary with no consistent direction, so its claimed set includes unsafe states, while ours starts from a conservative claim and enlarges it toward the true safe set without claiming unsafe states.}
\vspace{-15px}
\label{fig:teaser}
\end{figure}

Given dynamics and a safety constraint, Hamilton--Jacobi (HJ) reachability supplies exactly what a deployed robot needs, the maximal safe set together with a safety-preserving controller \citep{mitchell2005time,bansal2017hamilton}, but it supplies them for one obstacle configuration at a time.
When a warehouse robot faces a rearranged floor plan or a drone meets a new obstacle, the dynamics have not changed, yet the value function must be computed anew.
We focus on the cost of that recomputation.
Grid-based solvers re-solve a three-dimensional problem in tens to hundreds of milliseconds, but their cost grows exponentially with state dimension, which limits them to systems with no more than six state variables \citep{hsu2024safetyfilter}.
Self-supervised neural solvers of the DeepReach family \citep{bansal2021deepreach,singh2024exact,feng2025bridging} lift the dimension barrier, but they spend minutes to hours of training on each obstacle configuration, so a constraint that keeps changing puts them out of reach as well.
Classical remedies that avoid the re-solve, such as inflating obstacles by a precomputed tracking error bound \citep{herbert2017fastrack} or warm-starting a grid solution, pay in conservatism or remain grid-bound.
We therefore seek a single model of the map $c \mapsto V_c$ from obstacle configuration $c$ to value function $V_c$, trained once and queried in one forward pass.

A growing line of work learns exactly such conditioned models, from parameter-conditioned reachable sets \citep{borquez2023parameter} and neural operators over obstacle fields \citep{li2025hjrno,muenprasitivej2026contingency} to observation-conditioned value functions \citep{lin2024onefilter,derajic2025orncbf,feng2026gradientfree} and context-conditioned reachability deployed on a full-scale vehicle \citep{sahin2026context}, but three difficulties remain open across the line.
First, the model must cover every obstacle count, placement, and scene size that deployment can produce, a conditioning domain far too large to fill by sampling training scenes (the grid supervision that many of these models consume \citep{li2025hjrno,muenprasitivej2026contingency,lin2024onefilter,derajic2025orncbf} makes the domain costlier still, reinstating the curse of dimensionality at data-generation time).
Second, one network now carries a whole family of safety problems, so model error is unavoidable, and it falls on either side of the safety boundary with no consistent direction.
False-unsafe errors only cost performance, while false-safe errors are the failure a safety monitor exists to rule out, and most existing models accordingly describe their outputs as uncertified.
Third, a learned model can be trusted only where training data covered its input space, and false-safe errors concentrate in the configurations that data rarely covers, crowded obstacle clusters and long horizons, yet existing pipelines still draw their training scenes at random.

We propose a window-conditioned reachability model that addresses the above three gaps.
First, we cap the obstacle margin at a constant, which leaves the safe set unchanged and makes the value independent of every obstacle beyond the distance the robot can cover within the remaining horizon plus the constant (Lemma~\ref{lem:local}), so a generator that reads only this local window, in the robot's body frame, trains on small scenes and transfers to scenes of any size and obstacle count with millisecond scene preparation.
Second, we train $W$ to maximize the safe volume it claims, subject to the HJ inequality that keeps $W$ below the true value (Theorem~\ref{thm:sub}), so the remaining error falls on the conservative side, a control that equality-residual training, as in DeepReach, does not provide (Figure~\ref{fig:teaser}).
Third, we mine counterexamples over the finite-dimensional window configuration space that locality exposes (Proposition~\ref{prop:identity}) and replay the worst violations into training, so coverage reaches exactly the rare configurations that random scenes leave thin.

Our contributions are: (i)~a conservative-by-design formulation, a structural value bound plus inequality-constrained volume maximization, with an analysis of why equality-residual training is optimistically biased for avoid problems; (ii)~an exact local window structure that lets one model train on small scenes, deploy to scenes of any size with zero per-scene computation, and mine counterexamples over the finite-dimensional window space it exposes; (iii)~an evaluation on twelve Dubins obstacle settings against six conditioned or compositional baselines, where ours alone passes the false-safe and collision audit within the window capacity and attains the highest calibrated recovery, with a 10D quadrotor experiment that carries the method to high dimension.

\section{Mathematical Background}
\label{sec:background}

We consider control-affine dynamics without disturbance.
The state is $x \in \mathcal{X} \subseteq \mathbb{R}^n$, the control is $u \in \mathcal{U}$ with $\mathcal{U}$ compact, the dynamics are $\dot{x} = f(x) + g(x)\,u$, written $f(x,u)$ for short, with $f$ and $g$ Lipschitz; $\xi^{u}_{x}(s)$ denotes the state reached from $x$ after time $s$ under the control signal $u(\cdot)$.
Safety is encoded by a Lipschitz margin function $\ell_c : \mathbb{R}^n \to \mathbb{R}$ indexed by the constraint $c$, and the failure set is $\mathcal{F}_c = \{x : \ell_c(x) < 0\}$.
With $\tau \in [0, T]$ defined as the \emph{remaining} time, the avoid value function is
\begin{equation}
V_c(x, \tau) \;=\; \max_{u(\cdot)} \; \min_{s \in [0, \tau]} \; \ell_c\big(\xi^{u}_{x}(s)\big),
\label{eq:value}
\end{equation}
and the safe set at horizon $\tau$ is $\safeset_c(\tau) = \{x : V_c(x,\tau) \ge 0\}$.
The value function $V_c$ is the unique viscosity solution of the HJ variational inequality \citep{mitchell2005time,fisac2015reach}
\begin{equation}
\min\big\{\, \ell_c(x) - V_c, \;\; H(x, \nabla_x V_c) - \partial_\tau V_c \,\big\} = 0,
\quad V_c(x, 0) = \ell_c(x),
\quad H(x,p) = \max_{u \in \mathcal{U}} \, p \cdot f(x,u).
\label{eq:vi}
\end{equation}
In words, the value records the worst margin the robot will meet over the next $\tau$ seconds under the best possible control, and a state is safe exactly when that worst margin is nonnegative.

Safety certificates in this paper are built from one classical sufficient condition.
Define the residual of a differentiable candidate $W$ as $r_W(x,\tau) = H(x, \nabla_x W) - \partial_\tau W$.
We call $W$ a \emph{subsolution} of \eqref{eq:vi} if it satisfies the following two conditions, the \emph{value bound} and the \emph{HJ inequality},
\begin{equation}
\underbrace{W(x,\tau) \le \ell_c(x) \;\;\text{for all } (x,\tau)}_{\text{value bound}}
\qquad\text{and}\qquad
\underbrace{r_W(x,\tau) \ge 0 \;\;\text{on } \{W \ge 0\}}_{\text{HJ inequality}}.
\label{eq:subsol}
\end{equation}

\begin{theorem}[Inner Approximation; \citealp{fisac2015reach,xue2020inner}]
\label{thm:sub}
If $W$ is a subsolution of \eqref{eq:vi}, then $\{x : W(x,\tau) \ge 0\} \subseteq \safeset_c(\tau)$ for every $\tau$, and the feedback $u^W(x,\tau) \in \arg\max_{u} \nabla_x W \cdot f(x,u)$ renders $\{W \ge 0\}$ invariant while keeping $\ell_c \ge 0$ along the way.
\end{theorem}

We use Theorem~\ref{thm:sub} as the target of our training and do not claim it as a property of the learned model;
Appendix~\ref{app:proofs} recalls the two-line trajectory argument.\footnote{We follow the terminal-value convention of \citet{fisac2015reach}, under which \eqref{eq:subsol} is the subsolution side and subsolutions lower-bound the value.
Under the opposite sign convention the same object is a viscosity supersolution, whose zero-sublevel sets are the inner approximations synthesized in \citet{xue2020inner}.}
In words, any function that never exceeds the margin and whose claimed region satisfies the HJ inequality can only \emph{under}-claim safety: everything it marks safe truly is.
Beyond the switching feedback $u^W$, the same $W$ can also filter a reference control through a quadratic program, as a control barrier-value function \citep{choi2021robust}, with no retraining (Appendix~\ref{app:cbvf}).
Subsolutions are also a classical synthesis target, not only a verification condition: convex duality characterizes the value as the pointwise supremum over subsolutions \citep{vinter1993convex}, and sum-of-squares programming synthesizes polynomial subsolutions whose zero-superlevel sets form converging inner approximations of backward reachable sets \citep{korda2013inner,xue2020inner}.
Section~\ref{sec:method} transplants this synthesis approach to a neural, finite-horizon, constraint-conditioned setting.

\section{Problem Formulation}
\label{sec:problem}

Unlike the classical setting of Section~\ref{sec:background}, where the margin function is fixed before the value function is computed, we assume the safety constraint arrives at deployment time and keeps changing.
The $f, g$ functions in the dynamics are known, with a closed-form Hamiltonian, and belong to the translation--rotation-equivariant class that obstacle-avoidance problems possess.
The constraint is a finite set of disks $c = \{(o_k, \rho_k)\}_k$ with centers $o_k$ and radii $\rho_k$, inducing the signed-distance margin $\ell_c(x) = \min_k \big( \lVert p(x) - o_k \rVert - \rho_k \big)$ with $p(x)$ the robot position.
A new configuration $c$ with a different disk count, positions, radii, and scene extent can arrive at every query, so no computation that scales with the scene, and no retraining is admissible at deployment.
Ground-truth value functions are available neither at training time nor at deployment.
Training may use the HJ residual only, and grid solutions enter evaluation exclusively.

\textbf{Problem 1 (Conditioned Conservative Reachability).}
Given $f$ and the family of disk constraints, the problem is to learn a single model $W(x, \tau; c)$, queried in one forward pass per $(x,\tau,c)$, such that the model possesses: (i)~\emph{conservativeness}, errors falling on the side $W \le V_c$ so that no state the model claims safe is truly unsafe; (ii)~\emph{recovery}, the claimed set $\{W \ge 0\}$ covering as much of $\safeset_c(\tau)$ as possible subject to (i); (iii)~\emph{scene independence}, preparation of a new constraint requiring no retraining and no computation that grows with the scene, and each query costing the same regardless of scene size and obstacle count.

We build requirement (i) into the model and the objective, because requirement (iii) forbids per-scene verification or calibration.
Section~\ref{sec:experiments} instantiates the problem on a Dubins car, where grid ground truth makes requirement (i) measurable, and on a 10D quadrotor, where grids are unavailable and requirement (iii) is the only route left.

\section{Method}
\label{sec:method}

\begin{figure}[t]
\centering
\definecolor{figAcc}{HTML}{1F4E79}
\definecolor{figSnd}{HTML}{2F6B3A}
\definecolor{figWarn}{HTML}{B35C00}
\definecolor{figBad}{HTML}{A61B1B}
\begin{tikzpicture}[font=\sffamily\scriptsize,
  box/.style={draw=figAcc!70, rounded corners=2pt, fill=white, align=flush center, inner sep=2.5pt},
  gbox/.style={draw=figSnd!75, rounded corners=2pt, fill=figSnd!7, align=flush center, inner sep=2.5pt},
  obox/.style={draw=figWarn!75, rounded corners=2pt, fill=white, align=flush center, inner sep=2.5pt},
  arr/.style={-{Stealth[length=1.8mm]}, semithick, draw=gray!60!black},
  lab/.style={font=\sffamily\tiny, color=gray!30!black},
  lanelab/.style={font=\sffamily\scriptsize\bfseries, rotate=90, anchor=center}]
\fill[rounded corners=3pt, figAcc!6]  (0,0)     rectangle (15.2,-2.55);
\fill[rounded corners=3pt, figSnd!6]  (0,-2.7)  rectangle (15.2,-5.85);
\fill[rounded corners=3pt, figWarn!7] (0,-6.0)  rectangle (15.2,-7.6);
\node[lanelab, color=figAcc]           at (0.22,-1.275) {Query};
\node[lanelab, color=figSnd]           at (0.22,-4.275) {Training};
\node[lanelab, color=figWarn!85!black] at (0.22,-6.8)   {Deployment};
\node[box, text width=1.7cm, minimum height=2.25cm] (scene) at (1.44,-1.27) {Scene $c$\\[1pt]\textbf{any size},\\\textbf{any number}\\of disks};
\draw[draw=figAcc!70, rounded corners=2pt, fill=white] (2.66,-0.145) rectangle (5.36,-2.395);
\node[align=flush center, inner sep=1pt] at (4.01,-0.47) {Exact local window\\(Lem.~\ref{lem:local})};
\begin{scope}[shift={(4.01,-1.39)}, scale=0.6]
  \fill[figAcc!12] (0,0) circle (0.9);
  \fill[figAcc!3]  (0,0) circle (0.8);
  \draw[figAcc!80, thick] (0,0) circle (0.8);
  \draw[figAcc!80, thick, dashed] (0,0) circle (0.9);
  \fill[gray!65] (0.42,0.30) circle (0.13);
  \fill[gray!65] (-0.30,-0.40) circle (0.11);
  \fill[gray!38] (0.766,-0.643) circle (0.15);
  \fill[gray!18] (-1.12,0.45) circle (0.12);
  \draw[-{Stealth[length=1.4mm]}, very thick, figBad] (0,0) -- (0.36,0);
  \fill[figBad] (0,0) circle (0.05);
\end{scope}
\node[inner sep=1pt, font=\sffamily\fontsize{6}{7}\selectfont] at (4.01,-2.16) {Disks within $vT{+}\delta$};
\node[box, text width=2.7cm, minimum height=2.25cm] (enc) at (7.08,-1.27) {Set encoder (\S\ref{sec:local})\\[3pt]embeds the in-window disks\\in the body frame and pools\\them into one feature};
\node[box, text width=1.9cm, minimum height=2.25cm] (trunk) at (9.69,-1.27) {Value network\\reads $\tau$, $\lsoft$\\$\downarrow$\\four heads\\$g_1,\dots,g_4$\\(App.~\ref{app:impl})};
\node[gbox, text width=3.7cm, minimum height=2.25cm] (out) at (12.78,-1.27) {Conservative value $W$ (Eq.~\eqref{eq:head})\\[3pt]$W_i=\lsoft-\tau\,\mathrm{softplus}(g_i)$\\$W=\max_i W_i\le\lsoft\le\ell_c$\\[2pt](App.~\ref{app:heads})};
\draw[arr] (scene.east) -- (2.66,-1.27);
\draw[arr] (5.36,-1.27) -- (enc.west);
\draw[arr] (enc) -- (trunk);
\draw[arr] (trunk) -- (out);
\node[box, text width=1.7cm, minimum height=2.4cm] (batch) at (1.44,-4.27) {Batches (\S\ref{sec:mining})\\[2pt]scenes, plus draws from the window configuration space $\winspace$};
\node[gbox, text width=7.5cm, minimum height=1.05cm] (loss) at (6.55,-3.42) {Maximize the claimed safe volume\\under the HJ inequality (Eq.~\eqref{eq:loss})\\[2pt]$-\mathbb{E}\big[\mathrm{softmax}_i W_i\big]+\rho\,\mathbb{E}_{i\in A}\big[\relu(\mu-r_i)^2\big]$};
\node[box, text width=3.7cm, minimum height=1.05cm] (res) at (12.78,-3.42) {HJ residual (Eq.~\eqref{eq:subsol})\\[1pt]the slack $r_i$ of the HJ\\inequality at each head};
\node[box, text width=7.5cm, minimum height=0.9cm] (rho) at (6.55,-4.8) {Penalty multiplier $\rho$ adapted to a target violation rate;\\equality warm start in the first half (App.~\ref{app:impl})};
\node[box, text width=3.7cm, minimum height=1.2cm] (mine) at (12.78,-4.86) {Counterexample mining (\S\ref{sec:mining})\\[1pt]scene residuals are window\\residuals (Prop.~\ref{prop:identity}), so sweep\\$\winspace$ for the worst violations};
\draw[arr] (batch.north) -- node[right, lab] {inputs} (scene.south);
\draw[arr] (out.south) -- node[right, lab] {$W_i$} (res.north);
\draw[arr] (res.west) -- (loss.east);
\draw[arr] (rho.north) -- (loss.south);
\draw[arr] (res.south) -- (mine.north);
\draw[arr] ([xshift=1.1cm]loss.north) -- node[right, lab] {update $\theta$} ([xshift=1.1cm]loss.north |- enc.south);
\draw[arr] (mine.west) -- (10.62,-4.86) -- (10.62,-5.72) -- node[above, lab, pos=0.5] {worst points, replayed} (1.44,-5.72) -- (batch.south);
\node[obox, text width=3.6cm, minimum height=1.0cm] (prep) at (2.45,-6.8) {A \textbf{new scene} needs only its\\obstacle merge list (\S\ref{sec:local}),\\\textbf{no retraining}};
\node[obox, text width=4.8cm, minimum height=1.0cm] (range) at (7.45,-6.8) {Range check (\S\ref{sec:mining})\\a window outside the trained ranges\\of $\winspace$ is flagged as uncovered};
\node[obox, text width=3.7cm, minimum height=1.0cm] (filt) at (12.78,-6.8) {Safety filter (Thm.~\ref{thm:sub})\\$u^W\in\arg\max_u \nabla_x W\cdot f$};
\draw[semithick, draw=gray!60!black, densely dashed] (out.east) -- (14.98,-1.27) -- (14.98,-5.925) -- (2.45,-5.925);
\draw[arr, densely dashed] (2.45,-5.925) -- (prep.north);
\draw[arr, densely dashed] (7.45,-5.925) -- (range.north);
\draw[arr, densely dashed] (12.78,-5.925) -- (filt.north);
\end{tikzpicture}
\vspace{-25px}
\caption{Method overview.
\textbf{Query} reads only the disks within radius $vT+\delta$ and returns a conservative value $W$ in one forward pass, whatever the scene size.
\textbf{Training} maximizes the claimed volume subject to the HJ inequality, mining counterexamples over the window configuration space $\winspace$.
\textbf{Deployment} needs only a new scene's obstacle merge list, without retraining.
The green boxes above carry conservativeness.}
\vspace{-15px}
\label{fig:arch}
\end{figure}
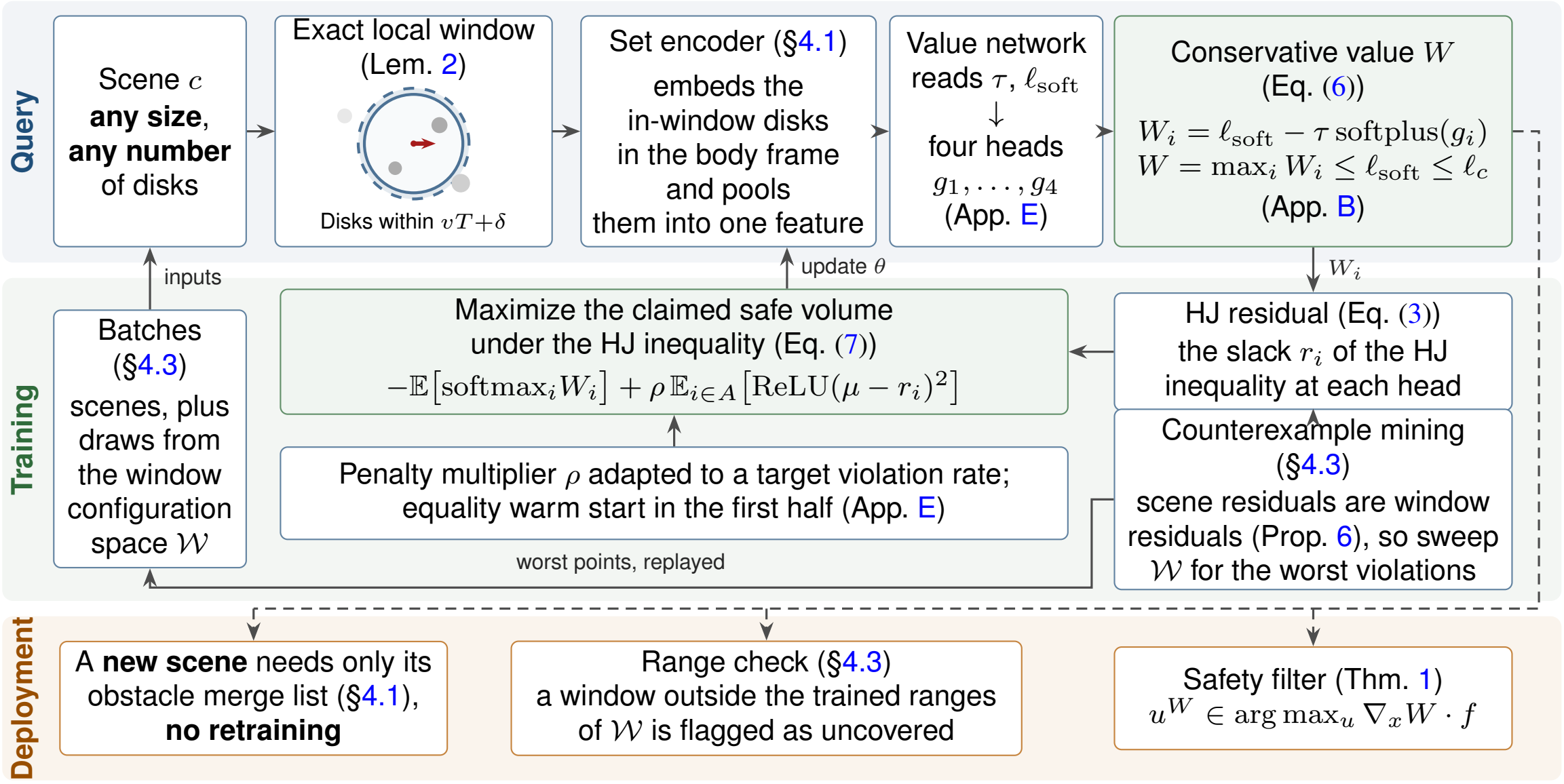

Our method reads, for each query, only the exact local window around the robot, encodes its disks in the body frame, and outputs a value capped by the margin; training maximizes the volume this value claims subject to the HJ inequality and mines counterexamples where the inequality fails.
Figure~\ref{fig:arch} summarizes the design, whose three components answer the three difficulties of Section~\ref{sec:intro}: an exactly local window carries constraint generality (Section~\ref{sec:local}), the output structure and the training objective carry conservativeness (Section~\ref{sec:objective}), and counterexample mining over the window configuration space carries coverage (Section~\ref{sec:mining}).

\subsection{Exact local windows: generalization from the domain of dependence}
\label{sec:local}

Our first observation is that capping the margin gives locality at no cost in the safe set.
We work throughout with the margin capped at a constant $\delta > 0$,
\begin{equation}
\lcap(x) = \min\big(\ell_c(x), \, \delta\big).
\label{eq:cap}
\end{equation}

\begin{lemma}[Margin Cap and Domain of Dependence]
\label{lem:local}
For any $\delta > 0$, the capped problem satisfies $V_{\lcap} = \min(V_c, \delta)$, so $\{V_{\lcap}(\cdot,\tau) \ge 0\} = \safeset_c(\tau)$ for every $\tau$.
Moreover, if the positional speed is bounded by $v$, then $V_{\lcap}(x, \tau)$ depends only on the disks $k$ with $d_k(x) \le v\tau + \delta$, where $d_k(x) = \lVert p(x) - o_k \rVert - \rho_k$.
\end{lemma}

The cap commutes with the pointwise minimum over trajectories, trajectories cannot travel farther than $v\tau$, and signed distances are 1-Lipschitz along positions.
In words, once the margin is capped, an obstacle farther than $v\tau + \delta$, and hence farther than the window radius $r_{\mathrm{in}} := vT + \delta$, provably cannot change the value.
We emphasize that this locality is exact for the capped value and that the window radius follows from the dynamics, whereas prior local-observation models fix their sensing ranges by hardware or design \citep{zhang2024gcbfplus,lin2024onefilter,sahin2026context}.
The full proof is in Appendix~\ref{app:proofs}.

We design the architecture to read exactly that window.
Disks within a slightly larger radius $r_{\mathrm{out}} > r_{\mathrm{in}}$ enter as body-frame features $(b_x, b_y, d_k, \rho_k)$, weighted by a smootherstep cutoff that equals $1$ inside $r_{\mathrm{in}}$ and $0$ at $r_{\mathrm{out}}$.
The window capacity is fixed in advance, six disks in this paper.
When more disks than the capacity intersect a window, pairs are merged into their smallest covering disk until the capacity is met; merging decreases $\ell_c$ pointwise, and the value is monotone in the margin, so merging can only err on the conservative side (Appendix~\ref{app:proofs}).
A permutation-invariant set encoder aggregates the per-disk embeddings $\phi_k$ three ways in parallel,
\begin{equation}
e = \Big[\, \textstyle\sum_k w_k\, \phi_k, \;\; \operatorname{attn}\big(\{\phi_k\}_k\big), \;\; \textstyle\sum_{k<l} \psi(\phi_k, \phi_l) \Big],
\label{eq:encoder}
\end{equation}
a sum weighted by the cutoff $w_k$, an attention readout over the nearest disks with a null token for empty windows, and a pooling over unordered disk pairs that exposes pairwise gaps.
The code $e$, together with $\tau$ and $\lsoft$ defined below, feeds a SIREN trunk \citep{sitzmann2020implicit} that outputs $K$ heads $g_1, \dots, g_K$, from which Section~\ref{sec:objective} builds the conservative values $W_1, \dots, W_K$ and takes their maximum.
Deployment inherits requirement (iii) directly: preparing a new scene costs only a lightweight merge of its obstacle list, and each query is one forward pass with value and spatial gradient, at a cost independent of the scene.
Layer sizes and the remaining implementation details are in Appendix~\ref{app:impl}, and ablations of the design choices in Appendix~\ref{app:ablation}.

\subsection{Conservative by construction and by objective}
\label{sec:objective}

We build conservativeness into the model twice, through the output structure and through the training objective.
Each head subtracts a nonnegative, time-scaled amount from a smoothed margin,
\begin{equation}
W_i(x, \tau; c) = \lsoft(x; c) - \tau \cdot \mathrm{softplus}\big(g_i(x,\tau;c)\big), \qquad i = 1, \dots, K,
\label{eq:head}
\end{equation}
where $\lsoft$ is the log-sum-exp soft minimum of the in-window signed distances, capped at $\delta$.
At $\tau = 0$ each head returns $\lsoft$, and for $\tau > 0$ the softplus can only pull it down, so the last layer builds the value bound of \eqref{eq:subsol} into the model up to the smoothing of the margin; this structure follows \citet{singh2024exact}, \citet{manda2024learning}, and \citet{derajic2025orncbf}.
We thus leave only the HJ inequality to training.

To return the largest safe set while keeping conservativeness, the model outputs the pointwise maximum of the heads, $W = \max_i W_i$.
The maximum preserves the structural bound $W \le \lsoft$ and requires the HJ inequality only at the maximizing head: where head $i^\star$ attains the maximum, $\nabla W = \nabla W_{i^\star}$ and hence $r_W = r_{i^\star}$, and at ties the directional derivatives of the maximum dominate those of each tied head (Appendix~\ref{app:heads}).
Note that the admissible set of \eqref{eq:subsol} is closed under maxima but not under averages (Appendix~\ref{app:heads}).

We train the model to claim as much as possible subject to the HJ inequality of \eqref{eq:subsol}, imposed as a penalty.
With collocation states $x$ sampled in scenes and $A(x,\tau)$ the set of \emph{active} heads, those within a small band of the maximum, we minimize
\begin{equation}
\mathcal{L} \;=\; -\,\mathbb{E}_{x,\tau}\Big[\operatorname{softmax}_{i} W_i(x,\tau)\Big]
\;+\; \rho \cdot \mathbb{E}_{(x,\tau),\, i \in A}\Big[ \relu\big(\mu - r_i(x,\tau)\big)^2 \Big],
\label{eq:loss}
\end{equation}
where $r_i = H(x, \nabla_x W_i) - \partial_\tau W_i$ uses the closed-form Hamiltonian, $\mu = 0.02$ is the \emph{required margin} on the HJ inequality, the volume term uses a log-sum-exp soft maximum so that all heads receive gradient, and the multiplier $\rho$ is adapted toward a target violation rate (Appendix~\ref{app:impl}).
The second term targets the HJ inequality $r_W \ge 0$ of \eqref{eq:subsol} by penalizing active heads whose residual falls below $\mu$, and under this inequality Theorem~\ref{thm:sub} keeps every claimed state inside the true safe set.
The first term pushes $W$ up, so the claimed set grows toward the true safe set from the conservative side.

Importantly, the next proposition shows that maximizing volume subject to \eqref{eq:subsol} still aims at the true value $V_c$ on the safe set.
\begin{proposition}[The Supremum of Subsolutions Is the Value]
\label{prop:sup}
For the avoid problem \eqref{eq:value} without disturbance, every $W$ satisfying \eqref{eq:subsol} obeys $W \le V_c$ on its claimed set $\{W \ge 0\}$, and $V_c(x,\tau) = \sup\{\, W(x,\tau) : W \text{ satisfies \eqref{eq:subsol}} \,\}$ wherever $V_c(x,\tau) \ge 0$.
Consequently, no subsolution claims an unsafe state, and every state with $V_c(x,\tau) > 0$ is claimed by some subsolution.
\end{proposition}
In words, the conditions in \eqref{eq:subsol} keep $W$ below $V_c$ where it claims safety and still let a sufficiently expressive $W$ approach $V_c$ on the safe set (proof sketch in Appendix~\ref{app:proofs}).

\vspace{1px} 
Learning a subsolution, which stays below $V_c$ on its claimed set (Proposition~\ref{prop:sup}), matters because training on the equality residual is biased toward the unsafe side, as shown by the next proposition.

\begin{proposition}[Equality Residuals Fold Optimistically]
\label{prop:fold}
For the avoid problem without disturbance, let $W$ be Lipschitz, satisfy the terminal condition $W(x,0) = \ell_c(x)$ of \eqref{eq:vi}, and satisfy the equality residual $\min\{\ell_c - W,\; r_W\} = 0$ almost everywhere.
Then $W \ge V_c$.
\end{proposition}

In short, a model $W$ that fits the equality residual almost everywhere lands on or above $V_c$, the side a safety certificate cannot tolerate, and equality-trained models do collide (Appendix~\ref{app:ablation}).
The full argument is  shown in Appendix~\ref{app:proofs}, and we close with the remark that motivates the one-sided penalty in \eqref{eq:loss}.

\begin{remark}[Optimism Is Visible under the Inequality]
\label{rem:asym}
If $W$ satisfies the value bound and, at some point, $W(x,\tau) - V_c(x,\tau) = b > 0$, then along the trajectory induced by $u^W$ from $(x,\tau)$ the integrated residual violation satisfies $\int \max(0, -r_W)\, ds \ge b$ (Appendix~\ref{app:proofs}).
A squared equality loss cancels positive against negative residuals and cannot see $b$; the one-sided penalty in \eqref{eq:loss} is charged at least $b$ for it.
\end{remark}

\subsection{Counterexample mining over the window configuration space}
\label{sec:mining}

Locality lets us collapse the conditioning domain to a finite-dimensional space that training can sweep exhaustively.
Counterexample-guided training usually sweeps the states of one fixed problem \citep{solanki2025care,zhang2024seev,yang2025scalable}, whereas our sweep spans state, horizon, and constraint at once.

\begin{proposition}[Scene Residuals Equal Window Residuals]
\label{prop:identity}
Let the dynamics be SE(2)-equivariant and the generator read only body-frame window features with the cutoff of Section~\ref{sec:local}.
Then for every scene $c$ and state $(x, \tau)$, both $W(x,\tau;c)$ and $r_W(x,\tau;c)$ equal their values at the canonical configuration $(\tau, \mathcal{N}_c(x)) \in \winspace$, where $\mathcal{N}_c(x)$ lists the in-window disks in the robot frame and $\winspace$, the \emph{window configuration space}, is the set of horizons paired with at most six body-frame disks.
\end{proposition}

In short, training sweeps $\winspace$ directly, with uniform draws in every batch and \emph{configuration mining} that adversarially replays the worst residual violations, so covering this one bounded space covers every scene, and a runtime check flags deployment queries whose windows leave the trained ranges of $\winspace$.
The full training and deployment loop is summarized in Algorithm~\ref{alg:summary} in Appendix~\ref{app:impl}.

\section{Experiments}
\label{sec:experiments}

We test our method on a 3D Dubins car (Section~\ref{sec:results}) for a detailed analysis in low dimension, and on a 10D quadrotor (Section~\ref{sec:quad}) to examine how the method extends to high dimension.
We measure recovery, the fraction of the true safe set that a model claims, and the collision rate, the fraction of closed-loop rollouts that start from claimed states, follow the model's own safety policy, and still hit an obstacle.
Recovery on the Dubins car uses the true HJ value from a grid solver, which also exposes false-safe states, claimed states that are truly unsafe beyond the grid tolerance.
On the quadrotor the true value cannot be computed, so recovery is measured against a reference safe set, the states from which at least one of about a thousand sampled open-loop control sequences avoids every obstacle, an inner bound of the true safe set.
Training protocol, setting definitions, baseline adaptations, per-setting tables, and the remaining figures appear in Appendices~\ref{app:impl}--\ref{app:extra}.
The statistical safety assurance of this and the quadrotor experiment, through the scenario-based level shift of \citet{lin2023generating}, is assessed in Appendix~\ref{app:extra}.

\begin{table}[h]
\centering
\footnotesize
\setlength{\tabcolsep}{3pt}
\caption{Dubins car, recovery at $\tau = 1$ in the nine obstacle settings within the window capacity and the largest rollout collision rate over them.
Big has a radius beyond the training range, far has its center outside the training area, and dead-end and large span $6.5\times$ the training area. $^{*}$: false-safe states in at least one scene of the setting; a $^{*}$ on the method name marks false-safe states in every scene of every setting. $^{\dagger}$: converted input.
All twelve settings are in Table~\ref{tab:test-recovery} (Appendix~\ref{app:extra}).}
\label{tab:test}
\begin{tabular}{l ccc ccc ccc c}
\toprule
& \multicolumn{3}{c}{One disk} & \multicolumn{3}{c}{Disk count} & \multicolumn{3}{c}{Layout} & Max coll. \\
\cmidrule(lr){2-4}\cmidrule(lr){5-7}\cmidrule(lr){8-10}
Method & nominal & big & far & 2 & 3 & 4 & small & dead-end & large & (\%) \\
\midrule
ORN-CBF$^{*\dagger}$ & 1.000 & 0.998 & 1.000 & 0.999 & 0.998 & 0.996 & 0.996 & 0.993 & 0.988 & 2.7 \\
OCR$^{*}$ & 0.999 & 0.997 & 1.000 & 0.998 & 0.997 & 0.995 & 0.996 & 0.992 & 0.994 & 1.1 \\
template & 0.997 & 0.992 & 0.998 & 0.995$^{*}$ & 0.991$^{*}$ & 0.989$^{*}$ & 0.988$^{*}$ & 0.988$^{*}$ & 0.985$^{*}$ & 1.1 \\
HJRNO$^{*}$ & 0.998 & 0.980 & 0.998 & 0.997 & 0.992 & 0.987 & 0.992 & 0.991$^{\dagger}$ & 0.995$^{\dagger}$ & 14.6 \\
PCR+ExactBC & 0.998 & 0.996 & 0.999 & 0.993$^{*}$ & 0.983$^{*}$ & 0.970$^{*}$ & 0.973$^{*}$ & 0.954$^{*\dagger}$ & 0.944$^{*\dagger}$ & 0.6 \\
certFNO & 0.904 & 0.795 & 0.951 & 0.844 & 0.773 & 0.717 & 0.690 & 0.728$^{*\dagger}$ & 0.593$^{*\dagger}$ & 1.2 \\
Ours & 0.997 & 0.994 & 0.999 & 0.995 & 0.990 & 0.985 & 0.988 & 0.967 & 0.980 & \underline{\textbf{0}} \\
\bottomrule
\end{tabular}
\end{table}
\vspace{-10px}

\begin{figure}[h]
\centering
\includegraphics[width=\linewidth]{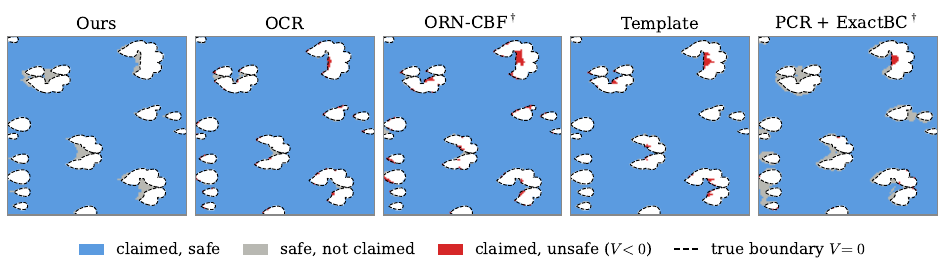}
\vspace{-25px}
\caption{Claimed sets on a dead-end scene at $\tau = 1$ and heading $0^\circ$.
Colors as in Figure~\ref{fig:teaser}. $^{\dagger}$: converted input.
The baselines claim unsafe states inside the dead-ends, where ours leaves safe states unclaimed.}
\vspace{-15px}
\label{fig:slices}
\end{figure}

\subsection{3D Dubins car}
\label{sec:results}

For the Dubins car, we test how safe a single trained model remains across twelve obstacle settings that vary the size, count, shape, and arrangement of the obstacles and the area of the scene, reaching well beyond the training range (definitions in Appendix~\ref{app:extra}).
When a window holds more than six obstacles, the model merges them, and the three \emph{crowded} settings where this happens, including the only failures of ours, are reported in Appendix~\ref{app:extra}, so Table~\ref{tab:test} covers the nine settings within the window capacity.
Baselines are six conditioned or compositional methods re-trained on our constraint family, ORN-CBF \citep{derajic2025orncbf}, OCR \citep{lin2024onefilter}, template composition \citep{solanki2026singleton} (Proposition~\ref{prop:minimum}), HJRNO \citep{li2025hjrno}, PCR+ExactBC \citep{borquez2023parameter,singh2024exact}, and certFNO \citep{muenprasitivej2026contingency}, plus trivial, grid-HJ, and per-scene DeepReach references.

Ours is the only method that is empirically safe in all nine settings, with no false-safe state, within the grid tolerance, and no collision (Table~\ref{tab:test}).
Every baseline produces false-safe states or collisions somewhere, and the failures grow with obstacle count.
OCR, ORN-CBF, and HJRNO produce false-safe states in every scene of every setting, with collision rates up to 1.1\%, 2.7\%, and 15\% respectively, and template composition and PCR+ExactBC fail in every setting with several obstacles.
certFNO stays free of false-safe states in seven settings, but only by surrendering recovery, 0.69--0.95 there.
A template variant that interpolates the single-disk values exactly reaches recovery 1.000 in all nine settings yet still misclassifies in every multi-disk setting and collides in 1.3\% of dead-end rollouts (Table~\ref{tab:test-recovery}).
This tells us that the bottleneck lies in the minimum composition itself, exactly as Proposition~\ref{prop:minimum} predicts, and the dead-end scenes are where it shows (Figure~\ref{fig:slices}).

Locality also lets our model handle scenes far larger than its training scenes.
Trained only on $[-1.21, 1.21]^2$ scenes, the model reaches recovery 0.980 in the large setting ($6.5\times$ area, 36 disks).
PCR+ExactBC, which reads world coordinates, keeps 0.66 there at matched statistical safety (Appendix~\ref{app:extra}).
Query time is scene-independent by construction.

Per-scene DeepReach needs 10--48\,min of training per scene to reach zero false-safe and, at matched statistical safety, the same disk-scene recovery as ours, rectangles excepted, while keeping 0.2--0.7\% rollout collisions, the optimistic fold of Proposition~\ref{prop:fold} in practice (Table~\ref{tab:per-scene}).
Ours trains once, prepares a new scene in 0.03--10\,ms against 53--248\,ms for grid re-solving, and answers 16{,}384 value-and-gradient queries in 2--4\,ms on one GPU, about 1\,ms per query on one CPU thread, which is fast enough for a 100\,Hz safety filter.

The component attributions and the equality-trained contrast are shown in Appendix~\ref{app:ablation}.

\begin{table}[h]
\centering
\small
\setlength{\tabcolsep}{4pt}
\caption{10D quadrotor on the scenes shared by all methods.
Rec.: recovery, the claimed fraction of the reference safe set.
Coll.: rollout collision rate in \%.
Full settings in Table~\ref{tab:quad-full} (Appendix~\ref{app:extra}). $^{\dagger}$: converted input (Appendix~\ref{app:baselines}).}
\label{tab:quad}
\begin{tabular}{l cc cc cc}
\toprule
& \multicolumn{2}{c}{Sparse} & \multicolumn{2}{c}{Dense} & \multicolumn{2}{c}{Large} \\
\cmidrule(lr){2-3}\cmidrule(lr){4-5}\cmidrule(lr){6-7}
Method & Rec. & Coll. & Rec. & Coll. & Rec. & Coll. \\
\midrule
PCR+ExactBC & 0.998 & 0.05 & 0.990$^{\dagger}$ & 0.60$^{\dagger}$ & 0.978$^{\dagger}$ & 0.18$^{\dagger}$ \\
Per-scene DeepReach & 1.000 & 0.10 & 0.997 & 0.29 & 0.998 & 13.2 \\
Ours & 0.995 & \underline{\textbf{0.004}} & 0.793$^{\dagger}$ & \underline{\textbf{0.16}}$^{\dagger}$ & 0.964$^{\dagger}$ & \underline{\textbf{0.025}}$^{\dagger}$ \\
\midrule
\emph{trivial} & 0.353 & - & 0.030 & - & 0.167 & - \\
\bottomrule
\end{tabular}
\end{table}
\vspace{-10px}

\begin{figure}[h]
\centering
\includegraphics[width=\linewidth]{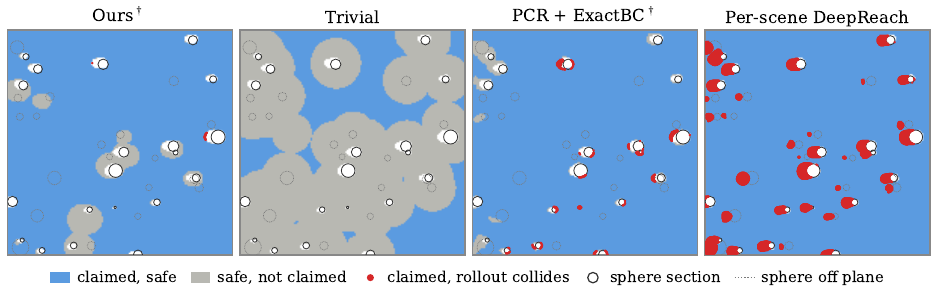}
\vspace{-25px}
\caption{Claimed sets on a scene of the large room ($24 \times 24$\,m) in the plane $z = 1.25$\,m for states moving at 1\,m/s along $+x$, at $\tau = 1$\,s. $^{\dagger}$: converted input.
PCR+ExactBC and per-scene DeepReach claim almost every safe state but also states from which their own safety policy collides, while ours leaves the regions around merged spheres unclaimed and collides only near the boundary of its claimed set.}
\vspace{-15px}
\label{fig:quad}
\end{figure}

\subsection{10D quadrotor}
\label{sec:quad}

For the quadrotor, we check how our method extends to high dimension.
First, a grid-based true HJ value is out of reach at this high dimension, so the baselines that need true values, HJRNO, certFNO, ORN-CBF, OCR, and the grid-based template composition, drop out.
The trivial reference, which claims no state that can reach an obstacle and thus never collides, is included only to gauge the difficulty of each setting.
We then test the quadrotor among spheres between a floor and a ceiling in three settings, sparse and dense obstacles in rooms of the training size and a room of nine times the area (model and protocol in Appendix~\ref{app:extra}).
Ours collides least in every setting (Table~\ref{tab:quad}, Figure~\ref{fig:quad}): 0.004--0.16\% of rollouts against 0.05--0.60\% for PCR+ExactBC and 0.10--13.2\% for per-scene DeepReach.
It recovers 0.995 and 0.964 of the reference safe set with sparse obstacles and in the large room but 0.79 with dense obstacles, where merged spheres exceed the training range in 59\% of windows.
Where a scene leaves the training range, both baselines keep their claims and pay in collisions, while ours answers windows it has not seen with conservativeness, giving up recovery rather than safety.

\section{Limitations and Conclusion}
\label{sec:limitations}

In this paper, we propose a conditioned reachability model that learns safe sets from the conservative side in order to avoid false-safe states, and we validate our method and compare against existing benchmarks on a Dubins car example and on a 10D quadrotor. However, there are still three limitations that bound the claims:
(a) No formal soundness is asserted.
The value bound in \eqref{eq:subsol} is built into the output structure, while the HJ inequality is a trained target, verified empirically within grid tolerance (Section~\ref{sec:results});
(b) Obstacles enter as disks, and other shapes and windows beyond six disks are covered and merged, which is conservative by monotonicity but pays recovery in the crowded settings;
(c) The local window restricts the method to avoid problems, since reach and reach--avoid specifications need global value information that a window cannot carry.
Within these bounds, one conditioned model serves scenes of any size at millisecond cost with its learning error directed to the safe side, and it is, to our knowledge, the first conditioned value function that survives a false-safe/collision audit within its window capacity while attaining the highest recovery after calibration.
In future work, we plan to verify the HJ inequality formally, handle general obstacle shapes, and extend the method to reach--avoid problems.

\subsection*{Generative AI Use}
Generative AI tools were used for literature search, coding and experimental implementation, and editing.
The authors reviewed and verified all AI-assisted content and take responsibility for the final content of the paper.

\newpage


\bibliography{refs}

\newpage
\appendix

\section{Proofs and Derivations}
\label{app:proofs}

\paragraph{Theorem~\ref{thm:sub}.}
Fix $(x, \tau)$ with $W(x,\tau) \ge 0$ and integrate $W$ along the closed-loop trajectory of $u^W$: $\frac{d}{ds} W(\xi(s), \tau - s) = \nabla_x W \cdot f(\xi, u^W) - \partial_\tau W = H(\xi, \nabla_x W) - \partial_\tau W = r_W \ge 0$ while $W \ge 0$, so $W$ never decreases along the trajectory, the trajectory stays in $\{W \ge 0\}$, and the value bound gives $\ell_c(\xi(s)) \ge W(\xi(s), \tau - s) \ge 0$ for all $s \le \tau$.
Hence $V_c(x,\tau) \ge 0$.
The statement is classical; see \citet{fisac2015reach} for the viscosity-solution formulation and \citet{xue2020inner} for the supersolution phrasing under the opposite sign convention.

\paragraph{Lemma~\ref{lem:local}.}
The map $m \mapsto \min(m, \delta)$ is nondecreasing and continuous, so it commutes with the pointwise minimum over $s$ in \eqref{eq:value} and with the supremum over controls, giving $V_{\lcap} = \min(V_c, \delta)$; since $\delta > 0$, the zero-superlevel sets agree.
For the dependence claim, a trajectory over horizon $\tau$ moves the position by at most $v\tau$, and each $d_k$ is 1-Lipschitz in position, so a disk with $d_k(x) > v\tau + \delta$ satisfies $d_k > \delta$ along the whole trajectory and never attains the capped minimum.

\paragraph{Monotonicity and merging.}
If $\ell \le \ell'$ pointwise then $V_{\ell} \le V_{\ell'}$, since the inequality holds along every trajectory and is preserved by $\min_s$ and $\sup_u$.
Replacing two disks by a covering disk decreases the margin pointwise, hence decreases the value: merging under-claims, never over-claims.

\paragraph{Proposition~\ref{prop:minimum}.}
$\ell_c = \min_k \ell_{c_k} \le \ell_{c_k}$ for each $k$, so by monotonicity $V_c \le V_{c_k}$ for each $k$ and $V_c \le \min_k V_{c_k}$.
If at $(x,\tau)$ some control avoids each disk individually but no control avoids all disks jointly, then $\min_k V_{c_k}(x,\tau) \ge 0 > V_c(x,\tau)$ and the inequality is strict.

\paragraph{Proposition~\ref{prop:identity}.}
Let $g \in \mathrm{SE}(2)$ map the robot state to the canonical pose (origin, zero heading).
Equivariance of $f$ makes the value and any function of body-frame features invariant under the simultaneous action of $g$ on state and obstacles; the cutoff zeroes every feature of disks outside the window, and Lemma~\ref{lem:local} shows those disks do not affect the capped value either.
Hence $W$ and, by the chain rule applied in the canonical frame, $r_W$ depend on $(x, \tau, c)$ only through $(\tau, \mathcal{N}_c(x))$.

\paragraph{Proposition~\ref{prop:sup}.}
Along the trajectory in the proof of Theorem~\ref{thm:sub}, $\ell_c(\xi(s)) \ge W(\xi(s), \tau - s) \ge W(x,\tau)$ whenever $W(x,\tau) \ge 0$, so $W \le V_c$ on the claimed set, and at a safe state the supremum is at most $V_c$.
For the reverse direction, fix $(x,\tau)$ and $\varepsilon > 0$, pick an $\varepsilon$-optimal piecewise-constant control $\bar{u}(\cdot)$, and define $\Phi(y, s) = \min_{r \in [0,s]} \ell_c(\xi^{\bar{u}}_{y}(r))$, the value of \emph{following} $\bar{u}$: $\Phi \le V_c$, $\Phi \le \ell_c$, $\Phi$ is nonincreasing along the $\bar{u}$-flow so its residual with respect to the linear Hamiltonian $p \cdot f(\cdot, \bar{u}(s))$ is nonnegative almost everywhere, and $H \ge p \cdot f(\cdot, \bar{u})$ pointwise, so mollifying $\Phi$ within each constancy interval yields a smooth subsolution within $O(\varepsilon)$ of $V_c(x,\tau)$ at the chosen point.
Convexity of $H$ in $p$ enters through Jensen's inequality in the direction that preserves the inequality for the linearized Hamiltonian.
This is the neural-setting analogue of the duality in \citet{vinter1993convex}; \citet{xue2020inner} prove the corresponding convergence for polynomial classes.

\paragraph{Proposition~\ref{prop:fold}.}
Write the equation in forward time $\tau$.
With control only, the Hamiltonian is convex in the gradient variable, and a Lipschitz function satisfying the equation almost everywhere is a viscosity supersolution of the forward-time equation by the convexity argument of \citet{barron1990semicontinuous}: at any point, the Clarke subdifferential is the convex hull of limits of gradients at points of differentiability, and convexity of $H$ transports the almost-everywhere inequality to every subdifferential element.
On the obstacle branch of the variational inequality the supersolution condition is automatic.
Comparison with the viscosity solution $V_c$, which satisfies the same terminal condition, yields $W \ge V_c$ under our sign convention.
In words, when a network fits the PDE except on a measure-zero set of kinks, the kinks resolve on the optimistic side.

\paragraph{Remark~\ref{rem:asym}.}
Suppose the value bound holds and $W(x,\tau) = V_c(x,\tau) + b$ with $b > 0$.
Run the trajectory of $u^W$ from $(x,\tau)$ and integrate as in the proof of Theorem~\ref{thm:sub}, but without assuming $r_W \ge 0$: $W(\xi(s), \tau - s) \ge W(x,\tau) - \int_0^{s} \max(0, -r_W)\, dr$.
If the integrated violation stayed below $b$, the trajectory would keep $W > V_c(x,\tau) - b + \mathrm{(violation)} \ge$ a margin forcing, by the value bound, $\min_s \ell_c(\xi(s)) > V_c(x,\tau)$, contradicting the definition of $V_c$ as the best achievable worst margin.
Hence $\int_0^{\tau} \max(0, -r_W)\, ds \ge b$.

\section{Multi-Head Details}
\label{app:heads}

\paragraph{Maximum preserves the structure.}
Each head satisfies $W_i \le \lsoft$, so $\max_i W_i \le \lsoft$ and the value bound holds for the combined output.

\paragraph{The inequality is needed only at the maximizing head.}
Where a single head attains the maximum, $W = W_{i^\star}$ in a neighborhood, so $\nabla W = \nabla W_{i^\star}$ and $r_W = r_{i^\star}$.
At ties, the lower Dini derivative of $\max_i W_i$ along any direction dominates the derivative of each tied head, so a nonnegative residual at every tied head implies the invariance argument of Theorem~\ref{thm:sub} goes through by following any tied head's feedback.

\paragraph{Averages are not admissible in general.}
With $H(x,\cdot)$ convex, $r_{\bar W} = H(x, \textstyle\sum_i \lambda_i \nabla W_i) - \sum_i \lambda_i \partial_\tau W_i \le \sum_i \lambda_i r_{W_i}$ is the wrong direction only when strict convexity bites: for Dubins, take two heads with $\partial_\tau W_i$ equal and $p_\theta$-components $+a$ and $-a$; each has residual contribution $1.1a$ from the $|p_\theta|$ term, the average has $0$, and a residual that was marginally nonnegative at each head becomes negative at the average.

\paragraph{Reporting policy.}
At inference the heads claim nearly identical sets; the maximum adds $0.03\%$ of grid cells beyond the best single head, so the benefit of $K = 4$ arises during training.
The paper reports the head-count ablation as a number and does not attach a mechanism to it.

\section{Using $W$ as a Control Barrier-Value Function}
\label{app:cbvf}

The feedback of Theorem~\ref{thm:sub} is least restrictive and ignores any nominal controller.
A $W$ satisfying \eqref{eq:subsol} can instead filter a reference control $u_{\mathrm{ref}}$ in the quadratic-program form of the control barrier-value function (CBVF) of \citet{choi2021robust}, with a rate $\gamma \ge 0$:
\begin{equation}
u^\star(x,\tau) \in \arg\min_{u \in \mathcal{U}} \; \lVert u - u_{\mathrm{ref}}(x,\tau) \rVert^2
\quad \text{s.t.}
\quad \nabla_x W(x,\tau) \cdot f(x,u) - \partial_\tau W(x,\tau) \;\ge\; -\gamma\, W(x,\tau).
\label{eq:cbvf}
\end{equation}
No $\gamma$-discounted training is needed, in contrast to the CBVF itself, which is the viscosity solution of a $\gamma$-discounted variational inequality.
Wherever $W \ge 0$, the HJ inequality gives $\max_{u} \nabla_x W \cdot f(x,u) - \partial_\tau W = r_W \ge 0 \ge -\gamma W$, so \eqref{eq:cbvf} is feasible with our learned $W$ as it stands.
For control-affine dynamics the constraint is linear in $u$, so \eqref{eq:cbvf} is a quadratic program.
Along the closed loop of any feasible control, $\frac{d}{ds} W(\xi(s), \tau - s) \ge -\gamma\, W(\xi(s), \tau - s)$, so Gr\"onwall's inequality gives $W(\xi(s), \tau - s) \ge e^{-\gamma s} W(x,\tau) \ge 0$, and the value bound keeps $\ell_c \ge 0$ along the trajectory, the argument of Theorem~\ref{thm:sub} with a graded intervention.
Small $\gamma$ intervenes earlier and more smoothly, and large $\gamma$ recovers the switching behavior.
In words, the learned value drops into a standard CBF-style filter unchanged, and the subsolution conditions are exactly what make the filter feasible on the claimed set.
As everywhere in this paper, the HJ inequality is the trained and empirically verified target, and the filter inherits that status.

\section{Ablation Studies}
\label{app:ablation}

All ablations are measured on five settings whose scenes are disjoint from those of Section~\ref{sec:experiments}, one to four disks and layouts that include dead-ends, on which Table~\ref{tab:validation} compares all methods (miss rate $= 1 -$ mean recovery at $\tau = 1$; seed noise $\approx 12\%$ relative).
Ablation numbers were measured with window capacity 8, while the final model is evaluated at capacity 6, which changes conditions only on the dead-end scenes.

\subsection{Training objective}
\label{app:abl-objective}

Raising the required margin from $\mu = 0.002$ to $0.02$ cuts the miss rate by 62\% (Table~\ref{tab:ablation}), with the adaptive multiplier settling at $\rho = 123$ instead of $8{,}650$.
The final recipe keeps the equality warm start: removing it and starting mining earlier is worse ($0.0127$, $+44\%$ over the final model, with dense-window recovery dropping to $0.902$), and the difference concentrates at $\tau = 1$ while $\tau = 0.5$ is unaffected.
The equality-only row quantifies the price of the inequality: it misses only $0.0069$ of the safe set but produces false-safe states ($4.0 \times 10^{-4}$ without tolerance) and a $0.14\%$ collision rate.
Conservatism costs about $0.002$ of recovery.

\subsection{Architecture}
\label{app:abl-arch}

Head count matters at equal training budget: four heads miss 0.89\% of the safe set against 1.54\% for a single head and 1.47\% for two (Table~\ref{tab:ablation}).
Removing the nearest-disk attention ($+38\%$) or shrinking the cap--window pair ($+29\%$) is worse; the SiLU embedding collapses dense-window recovery to $0.733$; and disk self-attention, despite the best dense-window recovery ($0.935$), fails the necessary condition outright (false-safe $1.1 \times 10^{-4}$, collisions $1.98\%$).
Of 18 inequality-trained runs, 17 satisfy zero false-safe and zero collisions on all five settings, so weakened models lose recovery rather than safety.
The single exception is the self-attention variant, and the only objective that fails is the equality loss.

\begin{table}[h]
\centering\small
\caption{Ablations on the five settings of this appendix.
Left: training objective.
Right: architecture (relative to the final model). $^{\times}$: fails the necessary condition (false-safe or collisions).}
\label{tab:ablation}
\begin{tabular}{lc@{\hskip 1.5em}lc}
\toprule
Objective variant & Miss rate & Architecture variant & Miss rate \\
\midrule
$\mu = 0.002$ (base recipe) & 0.0235 & final model & 0.0089 \\
$\mu = 0.02$ (final) & \textbf{0.0089} & $K = 1$ / $K = 2$ heads & 0.0154 / 0.0147 \\
no warm start, mining from 10\% & 0.0154 & no nearest-disk attention & 0.0122 \\
combined with $\mu = 0.02$ & 0.0127 & cap $0.3$, window $0.9/1.0$ & 0.0114 \\
equality loss only & 0.0069$^{\times}$ & SiLU embedding & 0.0539 \\
 & & disk self-attention & 0.0110$^{\times}$ \\
\bottomrule
\end{tabular}
\end{table}

\subsection{Counterexample mining}
\label{app:abl-mining}

Configuration mining is measured on 2 million held-out points of the window configuration space never used in training: the violation rate $P(r < 0)$ is $7.86 \times 10^{-4} \pm 1.03 \times 10^{-4}$ (three seeds) with mining on, $2.16 \times 10^{-3}$ with mining off, and $1.61 \times 10^{-3}$ when only the collocation count is increased to match.

\section{Implementation Details}
\label{app:impl}

\begin{algorithm2e}[h]
\DontPrintSemicolon
\caption{Training and deployment of the window-conditioned value}
\label{alg:summary}
\KwIn{dynamics with Hamiltonian $H$, cap $\delta$, required margin $\mu$, horizon $T$}
replay buffer $\mathcal{B} \leftarrow \emptyset$\;
\For{each training step}{
  sample scenes $c$ and states $(x, \tau)$, keeping the window configurations $(\tau, \mathcal{N}_c(x))$\;
  add direct draws $w \sim \mathrm{Unif}(\winspace)$ and counterexamples replayed from $\mathcal{B}$\;
  evaluate $W_1, \dots, W_K$ by \eqref{eq:head} and $r_i = H(x, \nabla_x W_i) - \partial_\tau W_i$ for $i \in A(x,\tau)$\;
  take a gradient step on \eqref{eq:loss}; raise $\rho$ when the violation rate exceeds its target, lower it otherwise\;
  \If{a mining round is due}{
    scan fresh $w \sim \mathrm{Unif}(\winspace)$, ascend the violation $\max(0, -r_W)$, and push the worst cases to $\mathcal{B}$\;
  }
}
\tcp{deployment: the one trained $W$ serves every scene}
merge the obstacle list of a new scene $c$ down to the window capacity\;
\For{each query $(x, \tau)$}{
  flag the query if $(\tau, \mathcal{N}_c(x))$ leaves the trained ranges of $\winspace$\;
  \Return $W(x, \tau; c)$ in one forward pass, with the safe control $u^W$ of Theorem~\ref{thm:sub} on demand\;
}
\end{algorithm2e}

The Dubins car moves at $v = 0.6$ with $|u| \le 1.1$ over the horizon $T = 1$, the constants of the multi-vehicle benchmarks of \citet{singh2024exact}, with closed-form Hamiltonian $H(x,p) = v(p_x \cos\theta + p_y \sin\theta) + 1.1\,|p_\theta|$ and margin cap $\delta = 0.2$.
Training scenes are drawn on $[-1.21, 1.21]^2$ with zero to six disks, radii from $U[0.08, 0.25]$ with probability $0.8$ and from $U[0.25, 0.5]$ with probability $0.2$, overlaps allowed; the training schedule appears in Table~\ref{tab:hparams}.
The objective of Eq.~\eqref{eq:loss} is applied after an equality warm start over the first half of the schedule.
The warm start is an initialization device that carries $W$ into the vicinity of $V_c$; the direction of the final error is set entirely by the inequality phase, and running the equality loss to the end instead produces collisions (Appendix~\ref{app:ablation}), consistent with Proposition~\ref{prop:fold}.
The deployment range check on window content costs 1.0\,ms (GPU) or 2.5\,ms (CPU) per 1{,}024 queries.

\begin{table}[h]
\centering\footnotesize
\caption{Hyperparameters of the final model.}
\label{tab:hparams}
\begin{tabular}{llll}
\toprule
margin cap $\delta$ & 0.2 & required margin $\mu$ & 0.02 \\
window radii $r_{\mathrm{in}} / r_{\mathrm{out}}$ & 0.8 / 0.9 & active-head band & 0.02 \\
window capacity & 6 disks & soft-min/-max temperature & 0.02 / 0.002 \\
heads $K$ & 4 & $\rho$ target violation rate & 0.1\% \\
parameters & 258{,}180 & $\rho$ update & $\times 1.5$ / $\div 1.2$, $[1, 10^5]$, start 100 \\
steps & 137{,}400 & batch & 32 scenes $\times$ 1{,}024 states \\
optimizer & Adam $10^{-4}$, cosine & $\tau$ curriculum & $0.1T \to T$ over first 20\% \\
equality warm start & first 50\% of steps & configuration-space scene share & 25\% of batch scenes \\
configuration mining & 4{,}096 scanned / 64 kept & state mining & residual-ascent perturbation \\
\bottomrule
\end{tabular}
\end{table}

Ground-truth values for evaluation come from a semi-Lagrangian grid solver ($\Delta x = 0.011$, 128 heading samples, $\Delta t = 0.05$) cross-validated three ways: grid refinement, closed-loop rollout consistency, and an independent reachability toolbox; the combined tolerance is $0.02$, used in the false-safe definition.

\section{Baselines and Fairness}
\label{app:baselines}

Each baseline is re-trained on our constraint family from its public recipe; grid-supervised methods receive grid solutions on our scenes as training data.
PCR+ExactBC conditions on the raw parameters of at most six disks; in the eight-disk setting and the three enlarged layouts (dead-end, large, cluttered) it reads the six disks nearest to the start state, kept fixed along each rollout, and in the rectangle setting the six nearest covering disks; ORN-CBF is evaluated through a tiling workaround in all settings; HJRNO and certFNO use grid-extension workarounds in the enlarged layouts.
Training-time comparisons carry two caveats: baselines trained on a shared GPU, ours on a slower Max-Q card, and PCR, ORN-CBF, and OCR trained for less than half our wall-clock.
Ours runs with three seeds, the baselines with one.
The trivial reference is $\min(\ell_c - v\tau,\, \delta)$, a zero-effort subsolution.
Template composition takes the pointwise minimum of single-disk values precomputed on a grid, the singleton composition that \citet{solanki2026singleton} present as a certificate of unavoidable collision and tighten with learned local refinement.
Our baseline uses the minimum alone and reads it as a safe set, as a safety filter requires, where it can only over-claim.

\begin{proposition}[Per-Obstacle Composition Is Optimistic; \citealp{solanki2026singleton}]
\label{prop:minimum}
Let $c = \bigcup_k c_k$ be a union of single-disk constraints.
Then $V_c \le \min_k V_{c_k}$, and the inequality is strict wherever no single control signal avoids all disks simultaneously.
\end{proposition}

The over-claim concentrates where obstacles interact, such as a dead-end formed by several disks; Section~\ref{sec:results} measures this failure directly, and the proof is in Appendix~\ref{app:proofs}.
The per-scene DeepReach reference uses exact terminal-condition imposition \citep{singh2024exact} at budgets of 6{,}000 / 24{,}000 / 120{,}000 iterations on two scenes per setting.

\section{Additional Results}
\label{app:extra}

\paragraph{Obstacle settings.}
Twelve settings probe distinct failure axes; each is listed with the internal code that labels the rows of the appendix figures.
One disk: \textbf{nominal} (P1-id, 64 scenes, in-distribution), \textbf{big} (P1-oodR, 16, radii beyond training), and \textbf{far} (P1-oodC, 16, centers beyond the training area).
Disk count: \textbf{2/3/4} (PB-k2/k3/k4, 32 each, exactly 2/3/4 disks) and \textbf{8} (PB-k8, 16).
Layout: \textbf{small} (L-small, 32, an area of the training size) and, on $6.5\times$ the training area, \textbf{dead-end} (L-trap, 8, arcs of nearly touching disks whose inner radius is below the Dubins turning radius), \textbf{large} (L-big, 16, 36 disks), and \textbf{cluttered} (L-dense, 8, 40 disks).
Shape: \textbf{rectangles} (S-rect, 32, rectangles converted to disk covers).
In three settings, 8, cluttered, and rectangles, windows hold more than six disks and are merged; we call these settings \emph{crowded}, and the main text reports the other nine.
Merging increases covered obstacle area (8: $10.5\% \to 17.3\%$; rectangles: $5.5\% \to 7.3\%$), the price of the capacity bound, paid in conservatism.

\begin{table}[h]
\centering\small
\caption{The five ablation settings of Appendix~\ref{app:ablation} (seed 0; one to four disks and layouts that include dead-ends).
Mean recovery at $\tau = 1$, worst false-safe rate, worst rollout collision rate, parameter count, and training time.
Baseline training times were measured on a shared GPU; ours on a slower Max-Q card.}
\label{tab:validation}
\begin{tabular}{lccccc}
\toprule
Method & Recovery $\uparrow$ & False-safe $\downarrow$ & Collision $\downarrow$ & Params & Train time \\
\midrule
ORN-CBF & 0.9972 & $1.2 \times 10^{-3}$ & 1.32\% & 4.0M & 734\,s \\
OCR & 0.9967 & $5.6 \times 10^{-4}$ & 0.71\% & 0.84M & 5{,}036\,s \\
template & 0.9924 & $4.9 \times 10^{-5}$ & 0.45\% & --- & --- \\
HJRNO & 0.9918 & $2.2 \times 10^{-2}$ & 6.85\% & 42M & 19{,}298\,s \\
PCR+ExactBC & 0.9823 & $1.1 \times 10^{-5}$ & 0.22\% & 0.80M & 3{,}214\,s \\
certFNO & 0.7818 & $1.0 \times 10^{-6}$ & 0.14\% & 42M & 19{,}298\,s \\
\midrule
Ours & 0.9911 & \textbf{0} & \textbf{0} & 0.26M & 5.2\,h \\
\bottomrule
\end{tabular}
\end{table}

\paragraph{Dubins car, per setting.}
Tables~\ref{tab:test-recovery}--\ref{tab:test-tau05} give the per-setting numbers behind Table~\ref{tab:test} and add the three crowded settings.
Rollouts use the first eight scenes of each setting and calibration the first four (four and two in large and cluttered), so raw and calibrated recovery cover different scenes.
PCR+ExactBC reads the six disks nearest to the start state in the eight-disk setting and the enlarged layouts and the six nearest covering disks in rectangles; recentering its inputs at the robot raises its calibrated recovery in the large setting from 0.66 to 0.74.
The grid HJ reference itself needs a calibration shift of 0.014--0.044 and collides in 0.09\% of rollouts, because its claimed set reaches the boundary, where grid and time discretization act; after calibration ours matches or exceeds it in the nine uncrowded settings.
Rollouts run for one second under the bang-bang policy $u = 1.1\,\mathrm{sign}(\partial_\theta W)$ from 2{,}000 claimed states per scene, and the calibration shift follows \citet{lin2023generating} with $\varepsilon = 10^{-3}$, $\beta = 10^{-16}$, and $N = 75{,}683$ rollouts.

\paragraph{Calibrated recovery.}
Raw recovery rewards optimism, so the fair comparison applies the same scenario-based level shift to every method, equalizing the statistical safety assurance, and compares recovery afterward.
Before calibration, OCR, ORN-CBF, and template sit above ours in most settings (ours: 0.967--0.999 in the uncrowded settings; 0.860, 0.879, 0.943 in the crowded ones, where merging charges its conservatism).
After calibration the ordering reverses: baseline recovery drops sharply, and more so the more disks a setting has (in cluttered, OCR falls to 0.72, template to 0.75, PCR+ExactBC to 0.61, and ORN-CBF to 0.44, while ours keeps 0.90).
Ours needs no calibration shift in eight settings and attains the highest calibrated recovery in every setting, tied with PCR+ExactBC in two single-disk settings (Table~\ref{tab:test-calibrated}).

\paragraph{Crowded settings.}
Merging costs recovery in the three crowded settings, where ours recovers 0.860 with eight disks, 0.879 in cluttered, and 0.943 on rectangles, and every failure of ours occurs here.
The false-safe states sit outside the trained configuration range, and a range check catches them.
The coverage analysis classifies every grid point by whether its window content lies inside the trained ranges of $\winspace$.
The 784 false-safe grid points of the single failing cluttered scene, $\sim 10^{-5}$ of the grid, all fall outside those ranges, at windows holding merged disks whose radius exceeds the training maximum of $0.5$, and \emph{no false-safe point lies inside them}.
The runtime range check of Section~\ref{sec:mining} flags exactly these inputs, so within declared coverage no false-safe state was observed, and outside it the check reports the input as uncovered rather than silently claiming safety.
The 2 collisions in 48{,}000 rollouts of the eight-disk setting start from states that the ground truth marks safe, so they are failures of the gradient policy rather than false-safe claims.

\begin{table}[h]
\centering
\scriptsize
\setlength{\tabcolsep}{1.2pt}
\caption{Dubins car: recovery at $\tau = 1$ per setting and the largest rollout collision rate over the settings. $^{*}$: false-safe states in at least one scene of the setting; a $^{*}$ on the method name marks false-safe states in every scene of every setting. $^{\dagger}$: converted input.
Rows in italics are references.}
\label{tab:test-recovery}
\begin{tabular}{l ccc cccc cccc c c}
\toprule
& \multicolumn{3}{c}{One disk} & \multicolumn{4}{c}{Disk count} & \multicolumn{4}{c}{Layout} & Shape & Max coll. \\
\cmidrule(lr){2-4}\cmidrule(lr){5-8}\cmidrule(lr){9-12}\cmidrule(lr){13-13}
Method & nominal & big & far & 2 & 3 & 4 & 8 & small & dead-end & large & cluttered & rectangles & (\%) \\
\midrule
HJRNO$^{*}$ & 0.998 & 0.980 & 0.998 & 0.997 & 0.992 & 0.987 & 0.955 & 0.992 & 0.991$^{\dagger}$ & 0.995$^{\dagger}$ & 0.989$^{\dagger}$ & 0.994 & 16.4 \\
certFNO & 0.904 & 0.795 & 0.951 & 0.844 & 0.773 & 0.717 & 0.570 & 0.690 & 0.728$^{*\dagger}$ & 0.593$^{*\dagger}$ & 0.547$^{*\dagger}$ & 0.813 & 1.2 \\
ORN-CBF$^{*\dagger}$ & 1.000 & 0.998 & 1.000 & 0.999 & 0.998 & 0.996 & 0.980 & 0.996 & 0.993 & 0.988 & 0.982 & 0.995 & 3.8 \\
OCR$^{*}$ & 0.999 & 0.997 & 1.000 & 0.998 & 0.997 & 0.995 & 0.986 & 0.996 & 0.992 & 0.994 & 0.992 & 0.997 & 1.3 \\
template & 0.997 & 0.992 & 0.998 & 0.995$^{*}$ & 0.991$^{*}$ & 0.989$^{*}$ & 0.979$^{*}$ & 0.988$^{*}$ & 0.988$^{*}$ & 0.985$^{*}$ & 0.983$^{*}$ & 0.976$^{*\dagger}$ & 1.1 \\
PCR+ExactBC & 0.998 & 0.996 & 0.999 & 0.993$^{*}$ & 0.983$^{*}$ & 0.970$^{*}$ & 0.898$^{*\dagger}$ & 0.973$^{*}$ & 0.954$^{*\dagger}$ & 0.944$^{*\dagger}$ & 0.929$^{*\dagger}$ & 0.950$^{*\dagger}$ & 0.6 \\
Ours & 0.997 & 0.994 & 0.999 & 0.995 & 0.990 & 0.985 & 0.860$^{\dagger}$ & 0.988 & 0.967 & 0.980 & 0.879$^{*\dagger}$ & 0.943$^{\dagger}$ & 0.004 \\
\midrule
\emph{trivial} & 0.687 & 0.486 & 0.844 & 0.535 & 0.421 & 0.337 & 0.157 & 0.304 & 0.303 & 0.159 & 0.163 & 0.503 & 0 \\
\emph{grid HJ} & 1.000 & 1.000 & 1.000 & 1.000 & 1.000 & 1.000 & 1.000 & 1.000 & 1.000 & 1.000 & 1.000 & 1.000 & 0.2 \\
\emph{template, exact interp.} & 1.000 & 1.000 & 1.000 & 1.000$^{*}$ & 1.000$^{*}$ & 1.000$^{*}$ & 1.000$^{*}$ & 1.000$^{*}$ & 1.000$^{*}$ & 1.000$^{*}$ & 1.000$^{*}$ & 0.983$^{*\dagger}$ & 1.3 \\
\bottomrule
\end{tabular}
\end{table}

\begin{table}[h]
\centering
\scriptsize
\setlength{\tabcolsep}{2.5pt}
\caption{Dubins car: calibrated recovery per setting.
Symbols as in Table~\ref{tab:test-recovery}.}
\label{tab:test-calibrated}
\begin{tabular}{l ccc cccc cccc c}
\toprule
& \multicolumn{3}{c}{One disk} & \multicolumn{4}{c}{Disk count} & \multicolumn{4}{c}{Layout} & Shape \\
\cmidrule(lr){2-4}\cmidrule(lr){5-8}\cmidrule(lr){9-12}\cmidrule(lr){13-13}
Method & nominal & big & far & 2 & 3 & 4 & 8 & small & dead-end & large & cluttered & rectangles \\
\midrule
HJRNO & 0.959 & 0.894 & 0.972 & 0.802 & 0.699 & 0.642 & 0.477 & 0.530 & 0.363$^{\dagger}$ & 0.194$^{\dagger}$ & 0.210$^{\dagger}$ & 0.759 \\
certFNO & 0.907 & 0.793 & 0.945 & 0.838 & 0.756 & 0.709 & 0.545 & 0.683 & 0.491$^{\dagger}$ & 0.266$^{\dagger}$ & 0.285$^{\dagger}$ & 0.781 \\
ORN-CBF$^{\dagger}$ & 0.992 & 0.968 & 0.996 & 0.978 & 0.919 & 0.860 & 0.639 & 0.840 & 0.734 & 0.496 & 0.438 & 0.877 \\
OCR & 0.988 & 0.955 & 0.992 & 0.978 & 0.922 & 0.908 & 0.720 & 0.898 & 0.797 & 0.789 & 0.718 & 0.905 \\
template & 0.996 & 0.984 & 0.997 & 0.991 & 0.976 & 0.910 & 0.810 & 0.936 & 0.734 & 0.844 & 0.750 & 0.799$^{\dagger}$ \\
PCR+ExactBC & 0.997 & 0.990 & 0.998 & 0.991 & 0.937 & 0.889 & 0.692$^{\dagger}$ & 0.878 & 0.771$^{\dagger}$ & 0.656$^{\dagger}$ & 0.609$^{\dagger}$ & 0.718$^{\dagger}$ \\
Ours & 0.997 & 0.995 & 0.998 & 0.996 & 0.991 & 0.984 & 0.837$^{\dagger}$ & 0.987 & 0.966 & 0.981 & 0.901$^{\dagger}$ & 0.929$^{\dagger}$ \\
\midrule
\emph{trivial} & 0.691 & 0.509 & 0.829 & 0.515 & 0.391 & 0.347 & 0.145 & 0.276 & 0.309 & 0.145 & 0.180 & 0.506 \\
\emph{grid HJ} & 0.996 & 0.985 & 0.997 & 0.994 & 0.979 & 0.968 & 0.907 & 0.972 & 0.953 & 0.952 & 0.939 & 0.970 \\
\emph{template, exact interp.} & 0.996 & 0.985 & 0.997 & 0.993 & 0.981 & 0.950 & 0.803 & 0.934 & 0.737 & 0.794 & 0.774 & 0.801$^{\dagger}$ \\
\bottomrule
\end{tabular}
\end{table}

\begin{table}[h]
\centering
\scriptsize
\setlength{\tabcolsep}{2.5pt}
\caption{Dubins car: recovery at $\tau = 0.5$ per setting. $^{\dagger}$: converted input.}
\label{tab:test-tau05}
\begin{tabular}{l ccc cccc cccc c}
\toprule
& \multicolumn{3}{c}{One disk} & \multicolumn{4}{c}{Disk count} & \multicolumn{4}{c}{Layout} & Shape \\
\cmidrule(lr){2-4}\cmidrule(lr){5-8}\cmidrule(lr){9-12}\cmidrule(lr){13-13}
Method & nominal & big & far & 2 & 3 & 4 & 8 & small & dead-end & large & cluttered & rectangles \\
\midrule
HJRNO & 0.996 & 0.984 & 0.998 & 0.996 & 0.991 & 0.986 & 0.960 & 0.991 & 0.991$^{\dagger}$ & 0.995$^{\dagger}$ & 0.991$^{\dagger}$ & 0.993 \\
certFNO & 0.907 & 0.805 & 0.953 & 0.852 & 0.786 & 0.735 & 0.595 & 0.709 & 0.747$^{\dagger}$ & 0.641$^{\dagger}$ & 0.599$^{\dagger}$ & 0.821 \\
ORN-CBF$^{\dagger}$ & 1.000 & 0.999 & 1.000 & 0.999 & 0.998 & 0.996 & 0.983 & 0.997 & 0.993 & 0.990 & 0.986 & 0.996 \\
OCR & 0.999 & 0.998 & 1.000 & 0.998 & 0.997 & 0.996 & 0.989 & 0.996 & 0.993 & 0.995 & 0.993 & 0.997 \\
template & 0.998 & 0.992 & 0.998 & 0.995 & 0.991 & 0.989 & 0.979 & 0.988 & 0.988 & 0.985 & 0.983 & 0.976$^{\dagger}$ \\
PCR+ExactBC & 0.999 & 0.998 & 0.999 & 0.994 & 0.988 & 0.982 & 0.951$^{\dagger}$ & 0.984 & 0.973$^{\dagger}$ & 0.972$^{\dagger}$ & 0.965$^{\dagger}$ & 0.965$^{\dagger}$ \\
Ours & 0.999 & 0.997 & 0.999 & 0.997 & 0.995 & 0.993 & 0.896$^{\dagger}$ & 0.994 & 0.988 & 0.991 & 0.898$^{\dagger}$ & 0.955$^{\dagger}$ \\
\midrule
\emph{trivial} & 0.896 & 0.784 & 0.948 & 0.828 & 0.759 & 0.703 & 0.558 & 0.659 & 0.688 & 0.562 & 0.538 & 0.797 \\
\emph{template, exact interp.} & 1.000 & 1.000 & 1.000 & 1.000 & 1.000 & 1.000 & 1.000 & 1.000 & 1.000 & 1.000 & 1.000 & 0.983$^{\dagger}$ \\
\bottomrule
\end{tabular}
\end{table}

\begin{table}[t]
\centering
\small
\setlength{\tabcolsep}{3.5pt}
\caption{Per-scene DeepReach with exact terminal conditions at three training budgets and ours, on two scenes from each of the settings nominal, 3, small, and rectangles.
Disks: mean over the three disk settings.
Training time is per scene for per-scene DeepReach. $^{\dagger}$: converted input.}
\label{tab:per-scene}
\begin{tabular}{l c cc cc c c}
\toprule
& & \multicolumn{2}{c}{Recovery} & \multicolumn{2}{c}{Calib.\ recovery} & False-safe & Max coll. \\
\cmidrule(lr){3-4}\cmidrule(lr){5-6}
Method & Training & Disks & Rectangles & Disks & Rectangles & scenes & (\%) \\
\midrule
Per-scene, 6{,}000 iters & 2.4 min & 0.995 & 0.979 & 0.979 & 0.929 & 7/8 & 0.4 \\
Per-scene, 24{,}000 iters & 9.7 min & 0.997 & 0.989 & 0.993 & 0.976 & 0/8 & 0.2 \\
Per-scene, 120{,}000 iters & 48 min & 1.000 & 0.995 & 0.995 & 0.988 & 0/8 & 0.7 \\
Ours & 5.2 h once & 0.994 & 0.929$^{\dagger}$ & 0.994 & 0.929$^{\dagger}$ & 0/8 & 0 \\
\bottomrule
\end{tabular}
\end{table}

\begin{figure}[t]
\centering
\includegraphics[width=\linewidth]{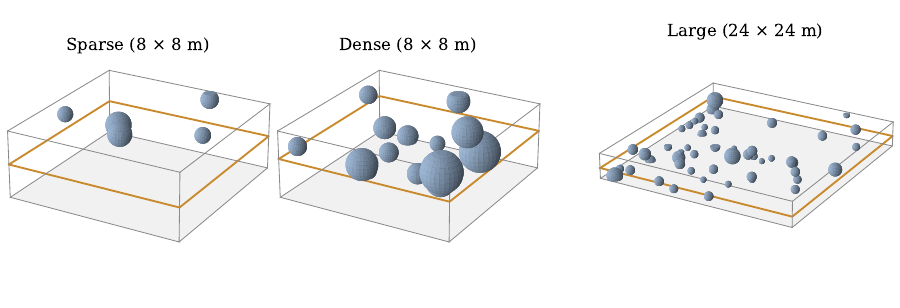}
\caption{The Sparse, Dense and Large scenes of the quadrotor slice figures, between the floor and the ceiling at 3\,m; the orange outline marks the plane of the slices.}
\label{fig:quad-envs}
\end{figure}

\paragraph{Quadrotor protocol.}
The tilt dynamics and their coefficients follow \citet{herbert2017fastrack}, the input sets differ from theirs, and the horizon $T = 1$\,s matches the quadrotor horizons of \citet{feng2025bridging,chen2018decomposition}.
The state $(p, v, \theta, \omega, z, w)$ stacks horizontal position, velocity, tilt and tilt rate in $\mathbb{R}^2$ with height $z$ and vertical velocity $w$; the inputs are a tilt command with $\|s\| \le 10^\circ$ and a vertical acceleration $|a| \le 1\,\mathrm{m/s^2}$, with gravity $9.81$, tilt coefficients $d_0 = 10$, $d_1 = 8$, $n_0 = 10$, and $\tan\theta \approx \theta$.
Obstacles are spheres of radius 0.3--1.0\,m, and the floor $z = 0$ and the ceiling $z = 3$\,m enter the margin.
The window radius, the largest displacement within the horizon plus the cap, is 2.5\,m.
Training scenes are $8 \times 8$\,m rooms with zero to six spheres; the Sparse setting (32 scenes) is drawn the same way, Dense (16) holds 10--14 spheres, and Large (16) is a $24 \times 24$\,m room with 41--59 spheres (Figure~\ref{fig:quad-envs}).
The reference safe set holds the states from which one of 1{,}027 open-loop controls (1{,}024 random piecewise-constant sequences and three heuristics) avoids every obstacle for one second, so it is an inner bound of the safe set.
Rollouts start from 2{,}000 claimed states in each of the first eight scenes (four on Large), and calibration uses the first four (two on Large).
Table~\ref{tab:quad} reports the scenes shared with per-scene DeepReach, four of Sparse, four of Dense and two of Large; Table~\ref{tab:quad-full} reports the full settings.
After the level shift, ours keeps the largest calibrated claim rate on Large, 0.78 against 0.69 for PCR+ExactBC and 0.04 for per-scene DeepReach.

\paragraph{Quadrotor violation target and safety checks.}
The quadrotor model lowers the violation target of the penalty schedule from 0.1\% to 0.01\% and keeps every other setting.
On separate scenes drawn like the training scenes, the fraction of claimed states with a negative residual falls from $1.0 \times 10^{-3}$ to $3$--$7 \times 10^{-5}$ across three seeds, and the claim rate drops by less than 0.01; the model trained with the Dubins target collides 2.5--15 times more often in the three settings (Table~\ref{tab:quad-full}).
In obstacle-free windows, where the exact value reduces to the vertical double integrator, none of $10^6$ sampled states is claimed unsafely at $\tau = T$.
From states within 0.5\,m of a sphere, ours collides in 0.3--0.5\% of rollouts, against 1.1--2.0\% for PCR+ExactBC.

\begin{table}[h]
\centering
\small
\setlength{\tabcolsep}{4pt}
\caption{10D quadrotor on the full settings.
Near coll.: collision rate of rollouts from claimed states within 0.5\,m of a sphere.
Ours (0.1\%) is the model trained with the Dubins violation target. $^{\dagger}$: converted input, merging for ours and the six nearest spheres for PCR+ExactBC.}
\label{tab:quad-full}
\begin{tabular}{l l cccc}
\toprule
Setting & Method & Recovery & Coll. (\%) & Near coll. (\%) & Calib.\ claim \\
\midrule
Sparse & trivial & 0.346 & 0 & -- & 0.346 \\
 & PCR+ExactBC & 0.998 & 0.11 & 1.34 & 0.915 \\
 & Ours & $0.996 \pm 0.0004$ & 0.010 & $0.3 \pm 0.1$ & $0.891 \pm 0.013$ \\
 & Ours (0.1\%) & 0.998 & 0.16 & 1.46 & 0.787 \\
\midrule
Dense & trivial & 0.065 & 0 & -- & 0.028 \\
 & PCR+ExactBC$^{\dagger}$ & 0.988 & 0.49 & 2.0 & 0.609 \\
 & Ours$^{\dagger}$ & $0.858 \pm 0.004$ & 0.13 & $0.5 \pm 0.3$ & $0.502 \pm 0.040$ \\
 & Ours (0.1\%)$^{\dagger}$ & 0.845 & 0.32 & 1.16 & 0.302 \\
\midrule
Large & trivial & 0.171 & 0 & -- & 0.169 \\
 & PCR+ExactBC$^{\dagger}$ & 0.980 & 0.16 & 1.1 & 0.685 \\
 & PCR+ExactBC, recentered$^{\dagger}$ & 0.995 & 0.23 & 1.8 & 0.747 \\
 & Ours$^{\dagger}$ & $0.969 \pm 0.001$ & 0.021 & $0.4 \pm 0.3$ & $0.781 \pm 0.020$ \\
 & Ours (0.1\%)$^{\dagger}$ & 0.969 & 0.18 & 1.55 & 0.590 \\
\bottomrule
\end{tabular}
\end{table}

\begin{figure}[p]
\centering
\includegraphics[width=\linewidth]{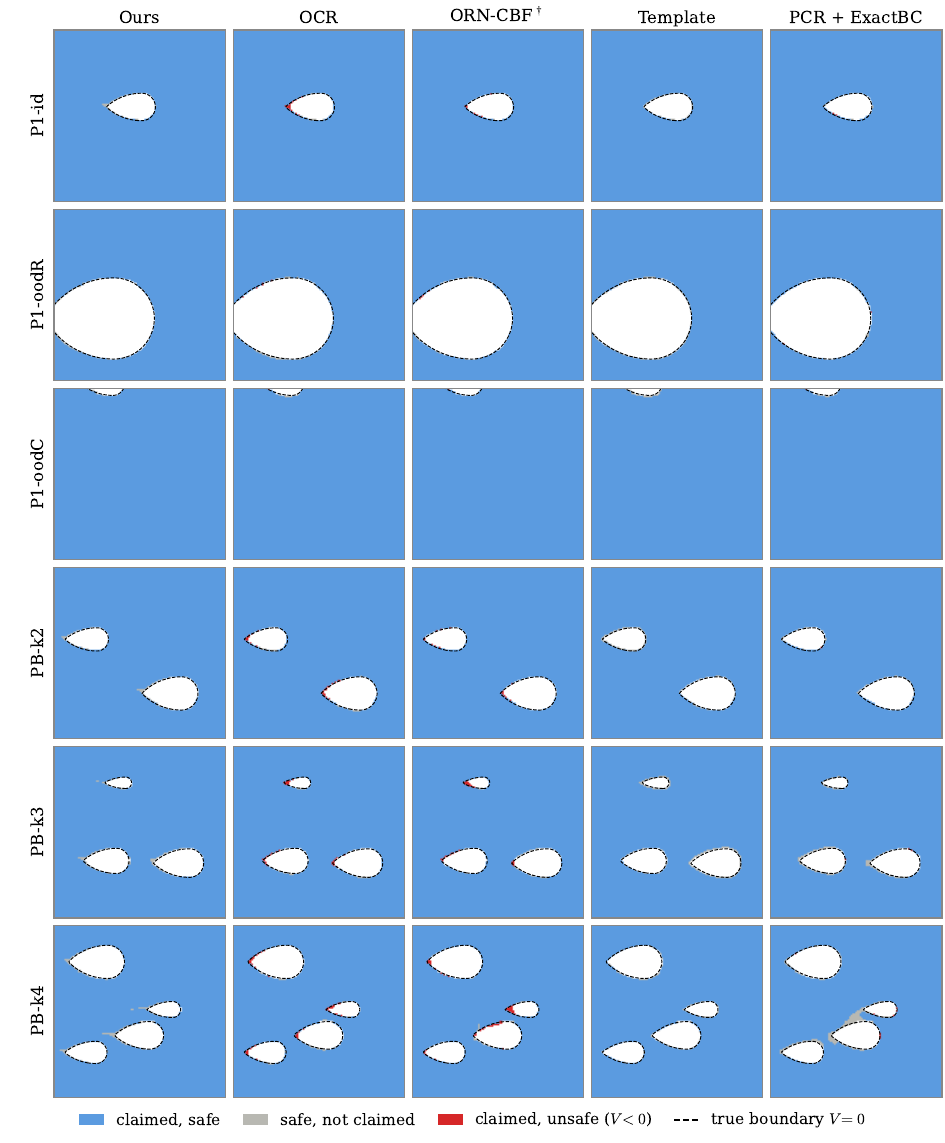}
\caption{Claimed sets at $\tau = 1$ and heading $0^\circ$ on the first scene of the one- to four-disk settings, drawn as in Figure~\ref{fig:slices}. $^{\dagger}$: converted input.}
\label{fig:slices-app-a}
\end{figure}

\begin{figure}[p]
\centering
\includegraphics[width=\linewidth]{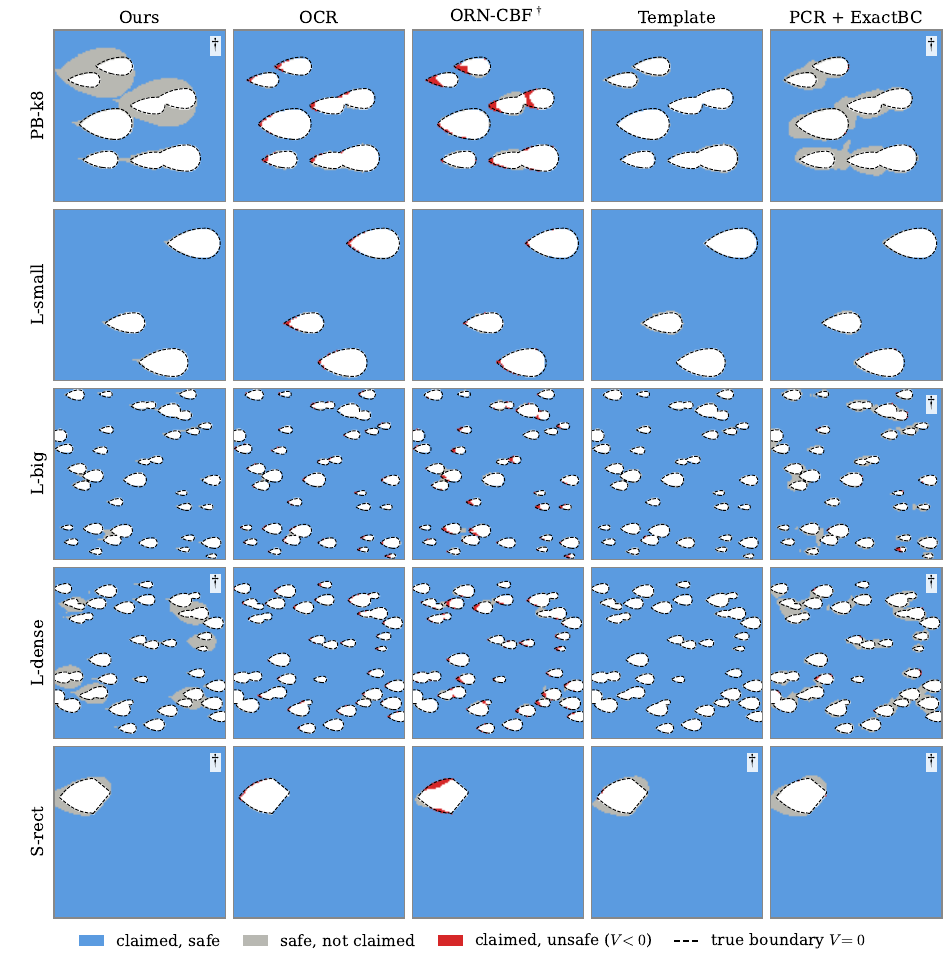}
\caption{Claimed sets at $\tau = 1$ and heading $0^\circ$ on the first scene of the settings 8, small, large, cluttered, and rectangles, drawn as in Figure~\ref{fig:slices}. $^{\dagger}$: converted input.}
\label{fig:slices-app-b}
\end{figure}

\begin{figure}[p]
\centering
\includegraphics[width=\linewidth,height=0.84\textheight,keepaspectratio]{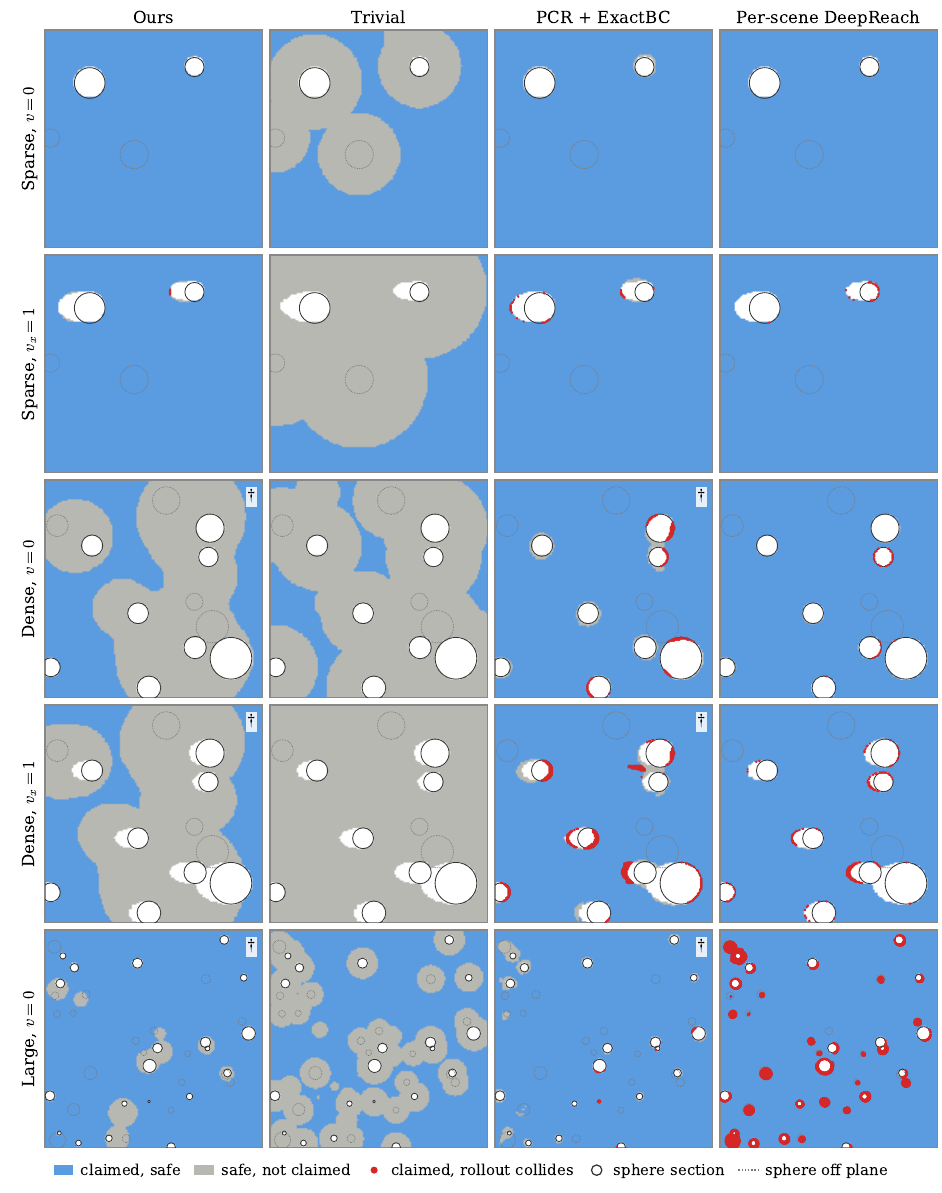}
\caption{Claimed sets on the Sparse and Dense scenes for hovering states and states moving at 1\,m/s along $+x$, and on the Large scene of Figure~\ref{fig:quad} for hovering states, in the planes $z = 1.5$, $1.75$ and $1.25$\,m at $\tau = 1$\,s, drawn as in Figure~\ref{fig:quad}. $^{\dagger}$: converted input.
On Dense, merged spheres remove much of our claimed set, the conservatism behind its lower recovery in Table~\ref{tab:quad}.}
\label{fig:quad-app}
\end{figure}

\begin{figure}[t]
\centering
\includegraphics[width=\linewidth]{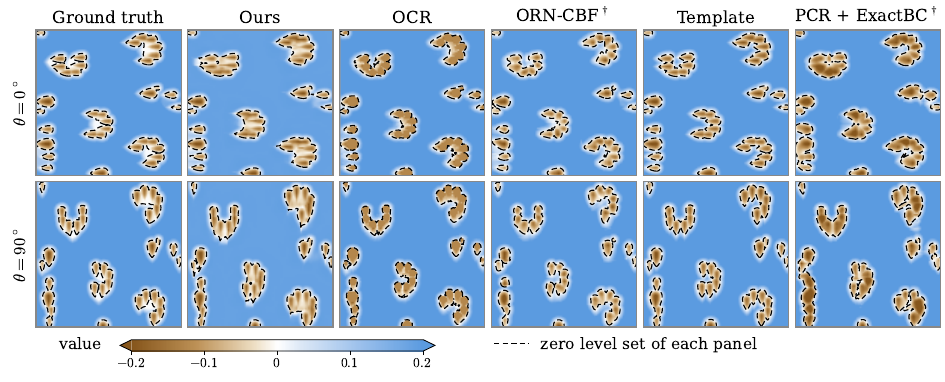}
\caption{Values on the dead-end scene of Figure~\ref{fig:slices} at $\tau = 1$ for headings $0^\circ$ (top) and $90^\circ$ (bottom). $^{\dagger}$: converted input.
Our value lies below the ground truth wherever it claims safety, up to the grid tolerance, so its zero level set stays inside the true one.}
\label{fig:heat}
\end{figure}

\begin{figure}[t]
\centering
\includegraphics[width=\linewidth]{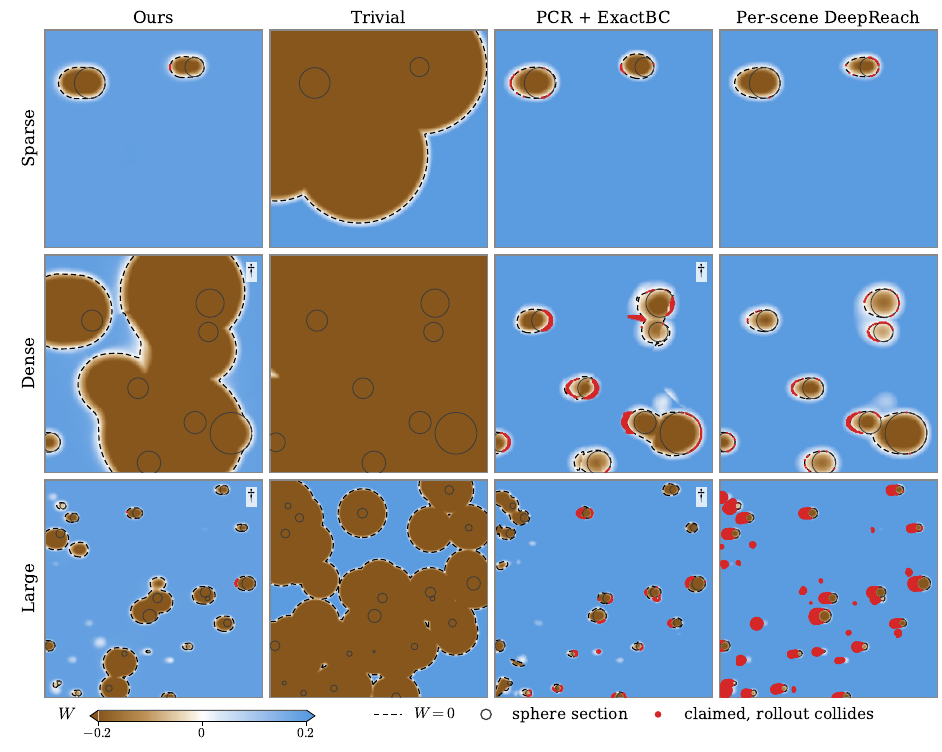}
\caption{Values $W$ on the quadrotor slices for states moving at 1\,m/s along $+x$, on the scenes of Figures~\ref{fig:quad} and~\ref{fig:quad-app}, with each panel's zero level set dashed. $^{\dagger}$: converted input.}
\label{fig:quad-heat}
\end{figure}

\paragraph{Value slices.}
Figures~\ref{fig:slices-app-a} and~\ref{fig:slices-app-b} show the claimed sets in the remaining Dubins settings, and Figure~\ref{fig:quad-app} shows those on the remaining quadrotor scenes; Figures~\ref{fig:heat} and~\ref{fig:quad-heat} show the values of both systems as heatmaps.
Each quadrotor slice uses the scene and height, among the scenes trained for per-scene DeepReach, whose plane cuts the most spheres.
On Large, per-scene DeepReach keeps the training budget of the $8 \times 8$\,m rooms on an area 8.6 times larger; all but 0.7\% of its claimed states lie in the reference safe set, yet its own safety policy collides from 13\% of them.

\end{document}